\documentclass[%
reprint,
superscriptaddress,
nobibnotes,
 amsmath,
 amssymb,
prl,
]{revtex4-2}

\usepackage[T1]{fontenc} 
\usepackage{amsmath}    
\usepackage{graphicx}   
\usepackage{verbatim}   
\usepackage{color}      
\usepackage{subfigure}  
\usepackage{hyperref}   
\usepackage{gensymb}
\usepackage{xcolor}
\begin{document}

\title{Noise Correlations in a 1D Silicon Spin Qubit Array}

\author{M.B. Donnelly}
\thanks{These authors contributed equally to the work}
\affiliation{Silicon Quantum Computing Pty. Ltd., Level 2 Newton Building, UNSW Sydney, 2052, NSW Australia}
\affiliation{School of Physics, UNSW Sydney, 2052, NSW Australia}
\affiliation{Centre for Quantum Computing and Communication Technology, UNSW Sydney, 2052, NSW Australia}

\author{J. Rowlands}
\thanks{These authors contributed equally to the work}
\affiliation{Silicon Quantum Computing Pty. Ltd., Level 2 Newton Building, UNSW Sydney, 2052, NSW Australia}
\affiliation{School of Physics, UNSW Sydney, 2052, NSW Australia}
\affiliation{Centre for Quantum Computing and Communication Technology, UNSW Sydney, 2052, NSW Australia}

\author{L. Kranz}
\affiliation{Silicon Quantum Computing Pty. Ltd., Level 2 Newton Building, UNSW Sydney, 2052, NSW Australia}
\affiliation{School of Physics, UNSW Sydney, 2052, NSW Australia}
\affiliation{Centre for Quantum Computing and Communication Technology, UNSW Sydney, 2052, NSW Australia}

\author{Y.L. Hsueh}
\affiliation{Silicon Quantum Computing Pty. Ltd., Level 2 Newton Building, UNSW Sydney, 2052, NSW Australia}
\affiliation{School of Physics, UNSW Sydney, 2052, NSW Australia}

\author{Y. Chung}
\affiliation{Silicon Quantum Computing Pty. Ltd., Level 2 Newton Building, UNSW Sydney, 2052, NSW Australia}
\affiliation{School of Physics, UNSW Sydney, 2052, NSW Australia}
\affiliation{Centre for Quantum Computing and Communication Technology, UNSW Sydney, 2052, NSW Australia}

\author{A.V. Timofeev}
\affiliation{Centre for Quantum Computing and Communication Technology, UNSW Sydney, 2052, NSW Australia}
\affiliation{School of Physics, UNSW Sydney, 2052, NSW Australia}

\author{H.\,Geng}
\affiliation{Silicon Quantum Computing Pty. Ltd., Level 2 Newton Building, UNSW Sydney, 2052, NSW Australia}
\affiliation{School of Physics, UNSW Sydney, 2052, NSW Australia}
\affiliation{Centre for Quantum Computing and Communication Technology, UNSW Sydney, 2052, NSW Australia}

\author{P. Singh-Gregory}
\affiliation{Silicon Quantum Computing Pty. Ltd., Level 2 Newton Building, UNSW Sydney, 2052, NSW Australia}

\author{S.K. Gorman}
\affiliation{Silicon Quantum Computing Pty. Ltd., Level 2 Newton Building, UNSW Sydney, 2052, NSW Australia}
\affiliation{School of Physics, UNSW Sydney, 2052, NSW Australia}
\affiliation{Centre for Quantum Computing and Communication Technology, UNSW Sydney, 2052, NSW Australia}

\author{\mbox{J.G. Keizer}}
\affiliation{Silicon Quantum Computing Pty. Ltd., Level 2 Newton Building, UNSW Sydney, 2052, NSW Australia}
\affiliation{School of Physics, UNSW Sydney, 2052, NSW Australia}
\affiliation{Centre for Quantum Computing and Communication Technology, UNSW Sydney, 2052, NSW Australia}

\author{R. Rahman}
\affiliation{Silicon Quantum Computing Pty. Ltd., Level 2 Newton Building, UNSW Sydney, 2052, NSW Australia}
\affiliation{School of Physics, UNSW Sydney, 2052, NSW Australia}

\author{M.Y. Simmons}
\email{michelle.simmons@unsw.edu.au}
\affiliation{Silicon Quantum Computing Pty. Ltd., Level 2 Newton Building, UNSW Sydney, 2052, NSW Australia}
\affiliation{School of Physics, UNSW Sydney, 2052, NSW Australia}
\affiliation{Centre for Quantum Computing and Communication Technology, UNSW Sydney, 2052, NSW Australia}

\date{\today}

\begin{abstract}
Correlated noise across multi-qubit architectures is known to be highly detrimental to the operation of error correcting codes and the long-term feasibility of quantum processors. The recent discovery of spatially dependent correlated noise in multi-qubit architectures of superconducting qubits arising from the impact of cosmic radiation and high-energy particles giving rise to quasiparticle poisoning within the substrate has led to intense investigations of mitigation strategies to address this. In contrast correlated noise in semiconductor spin qubits as a function of distance has not been reported to date. Here we report the magnitude, frequency and spatial dependence of noise correlations between four silicon quantum dot pairs as a function of inter-dot distance at frequencies from 0.3\,mHz to 1\,mHz. We find the magnitude of charge noise correlations, quantified by the  magnitude square coherence $C_{xy}$, are significantly suppressed from $>0.5$ to $<0.1$ as the inter-dot distance increases from 75\,nm to 300\,nm. Using an analytical model we confirm that, in contrast to superconducting qubits, the dominant source of correlated noise arises from low frequency charge noise from the presence of two level fluctuators (TLFs) at the native silicon-silicon dioxide surface. Knowing this, we conclude with an important and timely discussion of charge noise mitigation strategies.
\end{abstract}

\maketitle

\section{Introduction}

\begin{figure*}[]
    \centering
	\includegraphics[]{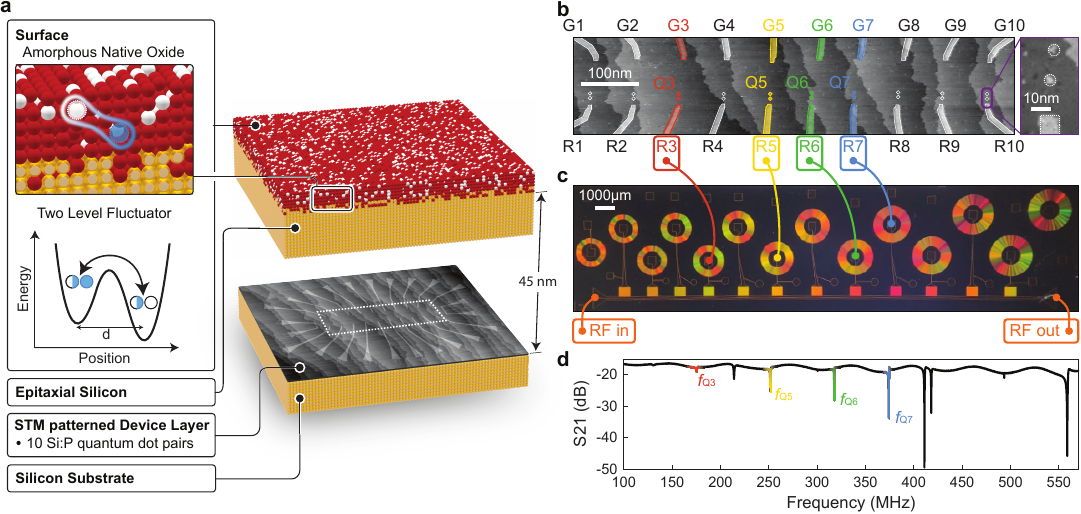}
	\caption{\textbf{Noise correlations in a 10 singlet-triplet multiplexed qubit array in silicon.} \textbf{a.} A diagram of the device material stack showing (from bottom to top): the silicon substrate, on which 10 Si:P quantum dot pairs containing $\sim$2 phosphorus atoms are patterned using STM lithography (see \textbf{b}). The quantum dots pairs are encapsulated with 45\,nm of epitaxial silicon which is terminated with an amorphous native silicon dioxide surface. Pairs of charge traps in the oxide form two level fluctuators (box, left) where electrons tunnel back and forth between the charge traps. This tunnelling forms a flip-flopping dipole $p=ed/2$, where $d$ is the separation between the traps, introducing correlated charge noise on the quantum dot pairs. \textbf{b.} STM image of the qubit array showing a linear set of ten pairs of quantum dots, each connected to a reservoir (R1-R10) and controlled by a gate (G1-G10). Four of the quantum dot pairs were studied in this work, marked Q3, Q5, Q6 and Q7, measuring charge noise PSD and correlations over inter-qubit separations from 75\,nm up to 300\,nm. \textcolor{black}{An STM image of a single Si:P quantum dot pair (Q10) is shown in the inset to the right showing two $\sim$2P dots separated by 11.4\,nm and 13.5\,nm away from the reservoir.} \textbf{c.} Each quantum dot pair reservoir is connected with a bond wire to an RF multiplexing chip containing NbTiN spiral inductors. Each inductor forms a separate LC resonant circuit for performing charge readout on each qubit individually. Simultaneous charge readout on multiple qubits is performed using frequency multiplexing through the single common RF transmission line (RF$_{in}$, RF$_{out}$). \textbf{d.} Frequency dependence of the S21 parameter of the multiplexing chip in transmission, with the relevant resonances for readout marked $f_{Q3}, f_{Q5}, f_{Q6}, f_{Q7}$.}
	\label{fig1}
\end{figure*}

Error correction remains a major challenge in realising a quantum processor that can perform calculations exponentially faster than classical computers. Quantum error correction requires that not only are errors sufficiently small but that they are uncorrelated \cite{Fowler2012, Fowler2014}, since errors found to be bunched in either space or time impede error correction protocols and significantly lower the threshold required for fault tolerance \cite{Clemens2004, Aharonov2006}. This poses a significant challenge for qubits in the solid state where phonons and electric field fluctuations in the qubit chip can lead to correlated qubit errors. For superconducting qubits, it has been recently shown that correlated errors can arise from the absorption of background cosmic radiation and high-energy particles within the substrate \cite{wilen2021, Martinis2021, Mcewen2022, google2021}. Such absorption events produce energetic phonons that break up the paired electrons in the superconducting material, inducing discrete changes in qubit offset charge via quasiparticle poisoning \cite{Patel2017, Liu2022, Mcewen2022}. These locally generated phonons propagate through the chip and induce spatial correlations in charge noise between qubits, typically observed on the $\sim$mm length scale \cite{wilen2021, Mcewen2022}. 

Substantial progress has been made in the past few years in another solid-state qubit platform - silicon spin qubits, with the recent demonstration of error rates below the fault tolerant threshold \cite{mkadzik2022, noiri2022, xue2022}. For quantum error correction (QEC) in this material system it is important that we also quantify and identify the source of correlated charge noise. Over the past decade, the magnitude of charge noise in silicon spin qubits primarily due to the presence of low frequency noise caused by two level fluctuators (TLFs) \cite{Connors2019, Freeman2016, zimmerman2014, Rudolph2019, takeda2013} has been systematically lowered by improving the materials quality and thereby mitigating the effect of these TLFs \cite{Connors2022, Kranz2020}. In particular, exceptionally low charge noise levels ($S_0=0.0088$\,$\mu$eV$^2$/Hz) have been observed in epitaxial spin qubits realised using phosphorus atoms embedded in high-quality crystalline silicon \cite{Kranz2020}.

In this work, we experimentally map the magnitude, frequency, and spatial dependence of noise correlations as a function of inter-qubit distance in a linear array of 10 phosphorus donor quantum dot pairs in a singlet-triplet qubit architecture with a separation of 75\,nm between neighbouring quantum dot pairs. Using frequency multiplexed charge readout we perform simultaneous multi-qubit charge noise measurements on 4 of the 10 dot pairs in the array using the technique of charge transition peak-tracking \cite{Kranz2020, Rudolph2019}. Here we simultaneously track the location in gate space of charge transitions from our charge sensors to several of the dot pairs and quantify noise correlations between them over a 16-hour period. Our results confirm a suppression of charge noise correlations as we increase the qubit separation up to 300\,nm. We also show a frequency dependence of the correlations which shows that they fall with increasing frequency up to 1\,mHz, a trend also observed in the SiGe material system \cite{Boter2020, Yoneda2022, rojasarias2023}. We then model the source of the charge noise correlations observed in our device by simulating charge traps at the native silicon/silicon-dioxide surface as TLFs.

\section{Results}

Charge noise correlations were measured in a phosphorus-doped silicon (Si:P) 20 dot singlet-triplet qubit \mbox{architecture \cite{Pakkiam2018}}, one of the largest silicon qubit architectures patterned using STM lithography to date (Figure \ref{fig1}a,b). Each singlet-triplet pair of phosphorus doped quantum dots is controlled by an electrostatic gate (G1-G10) and is tunnel coupled to a reservoir (R1-R10) with each reservoir connected to a dedicated resonant tank circuit on a separate chip (Figure \ref{fig1}c). All 10 tank circuits (consisting of superconducting thin film NbTiN spiral inductors) are connected via a common RF transmission line allowing for simultaneous frequency multiplexed readout of all quantum dot pairs  

\begin{figure}[]
	\includegraphics[]{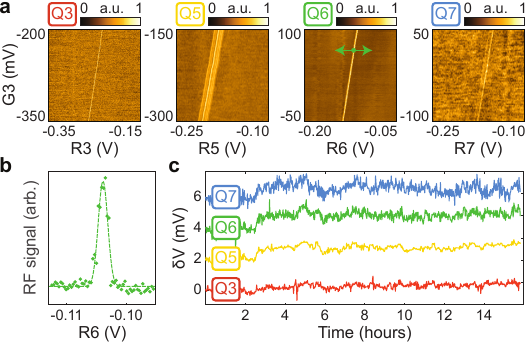}
	\caption{\textbf{Frequency multiplexed readout of four quantum dot pairs Q3, Q5, Q6 and Q7.} \textbf{a.} Measurements of the multiplexing chip RF response as the gate (G) and reservoir (R) voltages are swept for Q3, Q5, Q6, and Q7. The bright charge transition peaks running almost vertically indicate the loading of an electron from the reservoir onto the first dot, and are tracked over time to determine charge noise levels in the device. \textbf{b.} The RF response (circles) at the transition peak for Q6 (marked by arrows in \textbf{a.}) with a Gaussian fit (dotted line). The Gaussian fit is used to find the centre of the charge transition peak in voltage space. \textbf{c.} Simultaneous time traces of the location in voltage space of the reservoir to bottom dot charge transition for Q3, Q5, Q6 and Q7, with each data point taking 72\,s to measure and the entire trace measured over $\sim$16\,hours.}
	\label{fig2}
\end{figure}

To assess the spatially-dependent noise correlations in this device we considered four of the quantum dot pairs in the device, marked Q3, Q5, Q6 and Q7 as shown in Figure \ref{fig1}b. In Figure \ref{fig1}d we show the resonant frequency trace of the RF transmission line where the resonant frequencies of the tank circuits connected to Q3, Q5, Q6 and Q7 are highlighted. By probing the response of the transmission line at the resonant frequency of each tank circuit as we sweep the voltage applied to the reservoir $R$ and gate $G$, we can map out the charge stability diagrams for each quantum dot pair as shown in Figure \ref{fig2}a. To obtain the charge noise spectrum for each of the dot pairs we track the location of the reservoir to bottom dot charge transition for Q3, Q5, Q6 and Q7 simultaneously in the charge stability diagram over a period of $\sim$16\,hours \cite{Kranz2020}. The centre of the charge transition is obtained by fitting a Gaussian function to the peak in the RF signal (Figure \ref{fig2}b) with a single peak trace taking 72\,s. The measured peak positions of the four qubits Q3, Q5, Q6 and Q7 over the $\sim$16\,hour measurement are plotted in Figure \ref{fig2}c. From these time traces we calculated the charge noise power spectral density and interdot correlations.  

\begin{figure*}[]
	\includegraphics[]{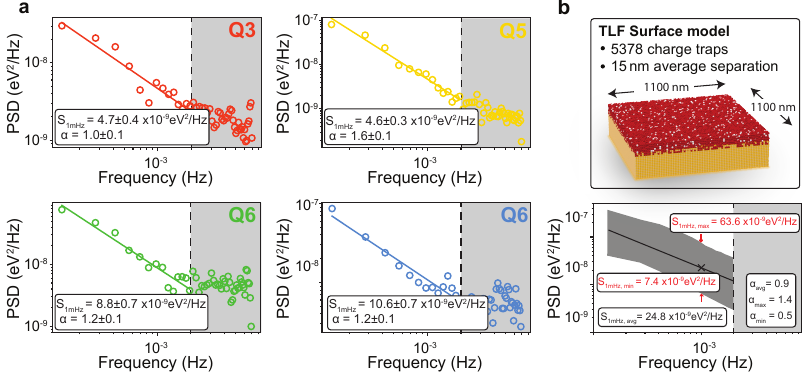}
	\caption{\textbf{Power spectral density of quantum dot pairs Q3, Q5, Q6, and Q7.} \textbf{a.} Measurements of the PSD for four quantum dot pairs in our array Q3, Q5, Q6, and Q7 (coloured circles) with fits to the data (solid line) used to extract the $S_{1mHz}$ and $\alpha$ parameters and uncertainties, before the onset of white noise at frequencies greater than 2\,mHz. The onset of white noise is indicated by the grey shaded region. \textbf{b.} (top) An illustration of the theoretical model used to simulate charge noise in the silicon qubit array. The surface oxide is populated with 5378 TLFs separated by an average distance of 15\,nm to model the charge traps that occur at the native oxide surface. (bottom) \textcolor{black}{Calculations of the PSD from the modelled time-dependent potentials $V(t)$ evaluated at over 500 points in the qubit plane. The average PSD over the >500 points is shown by the black line, and the range shown by the grey bar. The average modelled PSD amplitude at 1\,mHz is $S_{1mHz}=24.8\times10^{-9}$\,ev$^2$/Hz (black cross) with a range between $7.4\times10^{-9}$\,ev$^2$/Hz and $63.6\times10^{-9}$\,ev$^2$/Hz (red arrows). The average modelled spectral exponent is $\alpha$=0.9 with a range between 0.5 and 1.4.}
    }
	\label{fig3}
\end{figure*}

We first investigate the noise properties of each qubit individually by analysing the experimental time traces in Figure \ref{fig2}c and calculate the power spectral density (PSD), shown in Figure \ref{fig3}a. \textcolor{black}{We observe that the PSD of charge noise $S(f)$ follows a power-law dependence $S(f)=S_{1mHz}/(f\times10^{-3})^\alpha$ where $S_{1mHz}$ corresponds to the PSD value at 1\,mHz, and $\alpha$ determines the slope of $S(f)$ in frequency. Due to the focus on the low-frequency range of our experiment we compare the PSD amplitude at 1\,mHz $S_{1mHz}$, within the bandwidth of our experiment.} White noise is observed at frequencies $>$2mHz (grey shaded regions), where the device noise becomes lower than the experimental noise floor. The power-law nature of the low-frequency charge noise has been observed previously in semiconductor qubit platforms and attributed to the presence of an ensemble of TLFs; either formed at the surface, at dielectric interfaces, or within nearby oxides \cite{Shamim2016, Rudolph2019, Freeman2016, Connors2019, Yoneda2018, Connors2022, Kranz2020}. Due to their crystalline environment, donor devices have been shown to exhibit extremely low levels of charge noise since they are well separated from the surface TLFs by a protective layer of epitaxial silicon \citep{Shamim2016, Kranz2020}. \textcolor{black}{For the four quantum dot pairs measured in this device (see methods) we measure the PSD amplitude at 1\,mHz $S_{1mHz}$ with a range between $4.6\times10^{-9}$\,eV$^2$/Hz and $10.6\times10^{-9}$\,eV$^2$/Hz. For the spectral exponent we extract values in a range between 1.0 and 1.6. These values for $S_{1mHz}$ and $\alpha$ are consistent with previously reported values for donor devices \citep{He2019, Kranz2020, Koch2019}.}

To understand the nature and impact of the potential sources of noise in our devices we explicitly performed simulations of charge noise originating from TLFs located at the Si/SiO$_2$ native oxide surface located 45\,nm above the device. To model the switching noise of TLFs we created an analytical model (illustrated in Figure \ref{fig3}b, top) based on the standard tunnelling model of TLFs \cite{muller2019, martinis2005}. We populate an oxide layer (1100\,nm $\times$ 1100\,nm $\times$ 2\,nm) with charge traps separated by an average distance of 15\,nm as determined in previous native oxide surfaces by Shamim \textit{et. al.} \cite{Shamim2016}. This results in a total of 5378 charge traps populating the oxide interface. We consider that each trap consists of a positive charge $+e$ (where $e$ is the elementary charge of an electron), such that when pairs are formed between nearest-neighbour traps they form a double-well potential as shown in Figure \ref{fig1}a. Each pair shares two electrons so that the system remains charge neutral. The first electron in the pair gives a negative charge $-e$ bound by the two $+e$ traps giving a net charge of $e/2$ on each trap. When the second electron occupies only one of the traps, the trap pair becomes a dipole with a net charge $-e/2$ on the occupied trap and $+e/2$ on the unoccupied trap. As the second electron tunnels between the traps the dipole flip-flops creating a time-dependent potential with dipole moment $p=\pm ed/2$ where $d$ is the distance between the traps. This flip-flopping dipole generates a random telegraph signal (RTS) with a distribution given by $e^{-f_ct}$ where the centre frequency $f_c = f_0e^{-2\alpha_Td}$. Here $f_0$ represents the attempt frequency, typically on the order of $f_0 \sim 10^{12}$ Hz in solids \cite{Shamim2016} and $\alpha_T =\sqrt{2m^*\phi_B/\hbar^2}$, where $\phi_B=0.1$ eV is the conduction band offset for the native oxide trap states at the Si-SiO$_2$ interface \cite{Shamim2016} and $m^*$ is the electron effective mass. To generate the RTS in our model, we determine whether the dipole is flipping or not after a time interval $dt$ by picking a random number between 0 and 1, if the random number is higher than $e^{-f_c dt}$ then the dipole flips, otherwise the dipole remains the same. The fluctuating potentials from each flip-flopping dipole are then summed and evaluated at a depth of 45\,nm below the oxide layer at each qubit position according to $V(r)=\frac{1}{4\pi \epsilon_{Si}}\frac{p\cdot r}{r^3}$, where $\epsilon_{Si}$=11.7$\epsilon_{0}$ is the silicon dielectric constant, $\epsilon_{0}$ is the vacuum permittivity, and $r$ is the vector joining the dipole centre and the qubit location. The time-dependent potential sum $V(t)$ can then be compared directly to the peak position traces in Figure \ref{fig2}c measured on the device. \textcolor{black}{We evaluated the modelled time-dependent potential $V(t)$ at over $\sim$500 points evenly spaced over a 2D grid in the qubit layer with dimensions 300\,nm by 300\,nm. At each point we calculate the PSD of the modelled time-dependent potential $V(t)$, and in Figure \ref{fig3}b (bottom) we plot the average PSD in the simulated device (black line). Here the grey bar represents the variation in the PSDs across the device locations in the simulation with a average modelled PSD amplitude at 1\,mHz of $S_{1mHz}=24.8\times10^{-9}$\,ev$^2$/Hz and a range between $7.4\times10^{-9}$\,ev$^2$/Hz and $63.6\times10^{-9}$\,ev$^2$/Hz. Likewise, the average modelled spectral exponent is $\alpha$=0.9 with a range between 0.5 and 1.4. The ranges of the modelled values of $S_{1mHz}$ and $\alpha$ overlap with the our experimental values and are in agreement with previously reported values in Si:P \citep{He2019, Kranz2020, Koch2019}}

\begin{figure}[]
	\includegraphics[]{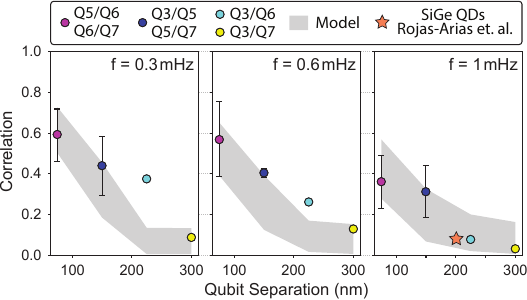}
	\caption{\textbf{Frequency dependence of spatial charge noise correlations.} \textcolor{black}{Measurements and simulations of the spatial charge noise correlations at frequencies of 0.3\,mHz, 0.6\,mHz and 1\,mHz. The coloured markers indicate correlations calculated using the experimental time traces. Two pairs of qubits we consider in our device are separated by 75\,nm and 150\,nm, and here we plot the average correlation value and use the difference to generate the error bars. The grey shaded region represents the average $\pm\sigma$ for 500 time traces generated using our model. At all frequencies we observe that correlations decrease with increasing qubit separation. At 1mHz we include data (orange star) from recent work in SiGe quantum dots \cite{rojasarias2023}.}}
	\label{fig4}
\end{figure}

We now turn to quantify the spatial dependence of charge noise correlations affecting the quantum dot pairs in the array. We use the magnitude-squared coherence function $C_{xy}(f)$, widely used in signal processing \cite{GARDNER1992113, MALEKPOUR20182932} to quantify the correlations between two time-dependent signals $x(t)$ and $y(t)$:

\begin{equation}
C_{xy}(f)=\frac{|S_{xy}(f)|^2}{S_{xx}(f)S_{yy}(f)},
\end{equation}

where $S_{xx}$ is the PSD of $x(t)$, $S_{yy}$ is the PSD of $y(t)$, and $S_{xy}$ is the cross-correlation PSD (cross-PSD) between $x(t)$ and $y(t)$. The magnitude-squared coherence $C_{xy}$ takes values between $C_{xy}$=0 (no correlation) and $C_{xy}$=1 (fully correlated) for the frequency components of $x(t)$ and $y(t)$. To measure the spatial dependence of the noise correlations we consider pairs of qubits according to the physical distance between them: Q5/Q6 and Q6/Q7 are separated by 75\,nm, Q3/Q5 and Q5/Q7 by 150\,nm, Q3/Q6 by 225\,nm and Q3/Q7 by 300\,nm. In Figure \ref{fig4} we plot $C_{xy}$ of the qubits pairs (taking the average of the two pairs of qubits at qubit separations of 75\,nm and 150\,nm) for three different frequencies (0.3\,mHz, 0.6\,mHz and 1.0\,mHz) spanning the accessible frequency range in our experiment. 
We used both experimentally measured time traces (shown by the coloured circles) and time traces generated from our analytical model (grey band). The correlation calculations from our theoretical model average the correlations over 500 time traces using the same trap distribution used in calculating the PSD. The grey regions in Figure \ref{fig4} represent the average of these 500 traces $\pm\sigma$. We observe that at 0.3\,mHz the modelled charge noise is highly correlated (0.5-0.7) between the qubits pairs separated by 75\,nm as confirmed experimentally where $C_{xy}$=0.6 was measured. At 0.3\,mHz the correlations drop to 0.1 as the separation increases to 300\,nm in both our modelling and experimental measurements. Here quantum dot pairs are less likely to be affected by the same TLFs as we increase the inter-dot distance. At higher frequencies (0.6\,mHz and 1\,mHz) we see a similar pattern where correlations fall as we increase the qubit separation. The correlations are, however, consistently smaller across all qubit separations at the higher frequency of 1\,mHz compared to 0.3\,mHz. This frequency dependence of noise correlations is consistent with recent studies in semiconductor quantum dots \cite{rojasarias2023, Boter2020}, where further work is needed to explore the relation between this frequency dependence and the microscopic origins of correlated noise. 

The results of our model are in agreement with our experimental data, capturing both the falling correlations with qubits separation and smaller correlations at higher frequencies. This provides evidence that the main sources of charge noise in our donor spin qubits are TLFs located at the native Si/SiO$_2$ interface. For comparison we also include a relevant data point from recent work performed in Si/SiGe quantum dots by Rojas-Arias \textit{et. al.} \cite{rojasarias2023} at 1mHz (orange star, Figure \ref{fig4}) where correlations were measured by observing fluctuations in the qubit energy. The separation between the qubit and the SiO$_2$ layer containing TLFs in the Si/SiGe device was 50\,nm, comparable to the 45\,nm in our Si:P device. Hence, we anticipate that future studies of charge noise correlations caused by TLFs in the SiGe-SiO$_2$ surface would have a similar spatial dependence to our results. Indeed, as observed in Figure \ref{fig4}, despite the different qubit spectroscopy methods involved, the magnitude of correlations from Rojas-Arias \textit{et. al.} is in good agreement with our experimental and modelling data. 

\section{Discussion}

Our results show that charge noise is also highly correlated in Si:P spin qubit architectures despite different origins for the superconducting qubit material system, and highlights the need for strategies to suppress the impact of correlated errors. The error correction threshold exponentially decreases as correlated errors couple to multiple qubits in a 2D array \cite{wilen2021}. It is therefore a priority to work in a regime where correlated errors can be effectively eliminated. Based on our measured noise correlations, four parallel routes towards a zero-correlation regime can be pursued: (1) increase the qubit separation distance to reduce the magnitude of correlated noise; (2) increase the frequency at which an error correction code can be run; (3) optimise the operating conditions to mitigate the effects of qubit errors; and (4) reduce the overall level of charge noise by optimising materials and device fabrication. In silicon spin qubits the qubit separation depends on the method of inter-qubit coupling, usually via exchange (10's of nm's to 100nm) or capacitive coupling (100's of nm's). Recent work in long range qubit coupling \cite{tosi2017, mi2017, wang2023} has been pursued primarily to make qubit architectures modular and more accessible for on-chip integration of control electronics. These advances may simultaneously benefit the suppression of correlated charge noise and its impact on qubit errors. The second route of increasing the frequency at which error correction codes can be run is achieved through pursuing faster qubit initialisation/readout \cite{keith2019} and control \cite{He2019}. Recent work has shown that noise correlations decrease with increasing frequency \cite{Boter2020} in a non-monotonic manner \cite{rojasarias2023}. While further work is needed to precisely map out noise correlations at higher frequencies than those studied here ($>$1\,mHz), we anticipate optimal error correction code operation frequencies exist where noise correlations are minimised. The third proposed solution involves implementing error mitigation techniques to reduce the impact of correlated errors on quantum computing algorithms. Such methods typically require extracting an accurate noise model of a quantum processor and carefully constructing the circuits such that the errors cancel themselves out \cite{strikis2021, temme2017, bravyi2021}. Whilst promising, these error mitigation techniques come with the cost of additional resource and/or time overheads. Finally, the overall impact of correlated noise can be minimised by removing the noise sources from qubit's proximity. Since our results suggest that the correlated noise originates from surface TLFs this can be suppressed in future devices by increasing the thickness of the encapsulation layer between the qubits and these surface oxide TLFs \cite{Shamim2016, Kranz2020}.

In conclusion, we have measured the power spectral density $S$ and magnitude squared coherence $C_{xy}$ in a linear array of ten quantum dot pairs using simultaneous charge state peak-tracking of four of the quantum dot pairs within the array. We find that correlations in the charge noise are suppressed by increasing the qubit separation (up to 300\,nm), as well as at the higher frequencies (up to 1\,mHz) studied in this work. Our experimental findings are supported by analytical modelling which considers the source of the charge noise as TLFs at the Si/SiO$_2$ interface. Our modelling replicates the trends observed in our experimental data and recent frequency dependence observed in Si/SiGe quantum dots \cite{rojasarias2023, Boter2020}. We observe that the spatial correlations in the Si:P qubit platform decay on the sub-$\mu$m length scale, compared to the $\sim$mm length scale found in superconducting qubits \cite{wilen2021}. These characteristic length scales can be attributed to the different underlying mechanisms causing the correlated noise: for silicon qubits the surface TLFs produce switching charge dipoles on the $\sim$10\,nm length scale, whereas for superconducting qubits phonons arising from the absorption of high-energy cosmic radiation leads to a larger spatial footprint of charging events \cite{wilen2021, Patel2017, Liu2022, Mcewen2022}. Phonons in silicon have been shown to affect relaxation and coherence times \cite{Hsueh2014, Hu2005}, as well as charge noise levels \cite{Petit2018}, however further work is necessary to determine their effect, if any, on charge noise correlations. Despite the relatively abrupt decay ($\sim$ 300 nm) of spatial correlations observed in our silicon samples, noise correlations pose significant challenges for implementing quantum error correcting codes. We discuss promising ways to mitigate these either by separating qubits by larger distances, increasing error-correction operation speeds, implementing recent error mitigation strategies \cite{strikis2021, temme2017, bravyi2021}, and/or improving device fabrication to minimise the known source of noise. In the short term, however, the knowledge gained about the spatially and temporally correlated noise in our semiconductor spin qubits can be leveraged in the NISQ era, where open quantum systems can be computationally powerful \cite{Aloisio2023} and noise correlations have been theoretically shown to improve the performance of variational quantum algorithms \cite{Kattemolle2023}.

\section{Acknowledgments}

The research was supported by Silicon Quantum Computing Pty Ltd and the Australian Research Council Centre of Excellence for Quantum Computation and Communication Technology (project number CE170100012). M.Y.S. acknowledges an Australian Research Council Laureate Fellowship. 

\section{Appendix: Materials and Methods}

\begin{table*}[]
\caption[]{\textbf{Summary of the key parameters of the multiplexing chip}. The resonant frequency $f_0$ and quality factors - total (internal, external), are calculated using S21 data. Gate leakage ranges are presented as between the Reservoir|Gate, and were measured out to 2\,V. Visible charge transitions are labelled as either reservoir to dot (RD) or interdot (ID).}
\begin{tabular}{|c|c|c|c|c|}
\hline
Qubit & $f_0$ (MHz) & $Q_t$ ($Q_i$, $Q_e$) & Leakage R|G & Transitions  \\
\hline
\hline
 1 & - & - & -0.7:0.9 | -2.0:2.0 & -  \\
 \hline
 2 & 129.8 & 130 (133, 1885) & -0.8:1.1 | -0.5:0.8 &  -\\
  \hline
 3 & 175.1 & 190 (227, 1096) & -0.5:1.2 | -1.0:1.2 & RD \\
  \hline
 4 & 213.9 & 273 (461, 661) & -0.7:1.1 | -2.0:2.0 &  -\\
  \hline
 5 & 251.7 & 293 (660, 421) & -0.8:1.0 | -0.9:1.1 & RD, ID \\
  \hline
 6 & 317.6 & 593 (1863, 789) & -0.6:0.9 | -0.6:1.0 & RD,ID \\
  \hline
 7 & 374.3 & 335 (1968, 330) & -0.6:0.7 | -0.6:1.0 & RD \\
  \hline
 8 & 414.2 & 362 (1753, 398) & -0.7:1.1 | -2.0:2.0 & RD \\
  \hline
 9 & 493.5 & 4246 (5071, 20416) & -0.7:1.1 | -1.1:1.9 & - \\
  \hline
 10 & 552.7 & 251 (1486, 302) & -0.9:1.2 | -2.0:2.0 & - \\
  \hline
\end{tabular}
\label{tab:table1}
\end{table*}

\subsection{Device Fabrication}

Fabrication of the 10 quantum dot pair device is performed using scanning tunnelling microscopy (STM) hydrogen resist lithography \cite{OBrien2001, Simmons2003, Shen1995, Lyding1994}, where an STM tip is used to selectively desorb a hydrogen mask bound to a 2$\times$1 reconstruction of the (001) silicon surface. This process is performed in an ultra-high vacuum (UHV) system with a base pressure of $\sim$1$\times$10$^{-11}$\,mbar. After lithography we dose the surface with phosphine gas (PH$_3$) at a pressure of $5\times10^{-7}$mbar for 2 minutes (1.8 Langmuir) and incorporate the phosphorus atoms into the exposed silicon using a 350\degree C anneal for 60 seconds. The device is encapsulated with a 45\,nm thick layer of epitaxial silicon grown at a temperature of 250\degree C and a growth rate of 0.15\,nm/min. To electrically contact the device once removed from the UHV environment, silicon vias aligned with STM-defined contact pads are etched using reactive ion etching and a paladium deposition and subsequent anneal is used to make Ohmic contact with the phosphorus-doped silicon layer. 

\subsection{Multiplexed RF Setup}

All measurements were performed in a dilution refrigerator with a base temperature of 10 mK. The 10 quantum dot pair chip was wire-bonded to a neighbouring multiplexed readout chip based on superconducting thin film NbTiN spiral resonators. The NbTiN spiral resonators have a nominal track width of 3\,$\mu$m and pitch of 3\,$\mu$m. The resonators connected to Q3/Q5/Q6/Q7 have 16/25/32/36 turns respectively with outer diameters of 795\,$\mu$m/903\,$\mu$m/987\,$\mu$m/1035\,$\mu$m. To provide multiplexed rf gate-based charge readout of the bottom dots tunnel-coupled to the reservoir leads (R1-R10), each resonator was coupled via an on-chip interdigitated capacitor to a common coplanar strip feed-line transmitting a multi-tone rf-signal. The resonators were driven with a nominal input power of $\approx$ -95\,dBm and the outgoing rf-signal was amplified by a 3 K stage low noise cryogenic pre-amplifier (CITLF2 Cosmic Microwave technology) with a nominal gain of $\approx$30 dB followed by two Pasternak HEMT amplifiers at room temperature with an overall 65 dB gain and the post-amplification chain.

The digital generation, control and demultiplexed readout of the high-purity multi-tone signals was implemented using a Keysight PXI M9019A FPGA-based chassis platform with M3201 AWG 500 MS/s DAC and 1 GS/s ADC M3102 digitiser modules. The generated frequency comb with an overall 200 MHz bandwidth was up-converted using a Polyphase AM0350A IQ-modulator and down-converted after the post-amplification chain with a Polyphase AD0105B IQ-demodulator. The signal was digitally demultiplexed using standard digital lock-in amplification and decimation techniques (a six state cascaded-integrator-comb filter and a decimation filter with a down-sampling factor $\times$3000). The voltage signal-to-noise ratios (SNR) for the charge transitions (correlated with intrinsic resonators' losses) were $\approx$10-20 at 1 kHz post-integration bandwidth being limited down to SNR$<2$ above 100 kHz bandwidth by the effective sampling rate of 390 kHz.

The DC-voltage bias of the reservoirs (R1-R10) was implemented via RC bias-tees, formed by the interdigitated capacitor (3 pF nominal) on the MUX chip and the PCB-mounted \mbox{1 M$\Omega$} MELF resistors wire-bonded to the DC-taps of the spirals.

\subsection{Device Characterisation}

\textcolor{black}{We calculated the charge noise correlations between sites Q3, Q5, Q6 and Q7. Of the 10 quantum dot paris fabricated these four had the correct combination of frequencies and tunnel rates required for high quality measurements, as shown in Table \ref{tab:table1}.}

\textcolor{black}{Here we summarise the key parameters of the multiplexing chip and the quantum dot pairs connected to them:}

\textcolor{black}{
\begin{itemize}
    \item Q1: No charge transitions were visible on this quantum dot pair as the resonant frequency $f_1$ was below the frequency floor of the demodulator (200\,MHz.)
    \item Q2: While reservoir to dot (RD) charge transitions were seen they were severely attenuated as the resonant frequency was still significantly below the frequency floor of the demodulator.
    \item Q3: RD charge transitions were seen at a resonant frequency $f_3$=175.1\,MHz by sweeping gates R3 and G3.
    \item Q4: There were no clear charge transitions in the gate space most likely due to the incorrect tunnel rate between the lower quantum dot and the reservoir. The resonant frequency used is $f_3$ = 213.9 MHz..
    \item Q5: RD charge transitions were seen at a resonant frequency $f_5$ = 251.7 MHz
    \item Q6: RD charge transitions were seen at a resonant frequency $f_6$ = 317.6 MHz. 
    \item Q7: RD charge transitions were seen at a resonant frequency $f_7$=374.3\,MHz.
    \item Q8: While three RD transitions were visible at a resonant frequency $f_8$=414.2 MHz, a small tip drop occurred during STM lithography between the quantum dots making the results incomparable.
    \item Q9: Whilst the bond wire to the multiplexing chip remained connected as evidenced by the order of magnitude higher quality factor compared to chip bonded resonators, the bond wire to the device became disconnected during cool down.
    \item Q10: Gate G10 was disconnected during cool down.
\end{itemize}
}
\textcolor{black}{
To avoid the device going into leakage (where current flows between control gates), we only scanned voltage ranges between -0.3V and 0.1V, remaining well within the leakage range of each gate (-0.8V to +1.1V). When the devices does go into leakage it must be warmed from mK to above the charge freeze-out temperature of the carriers which for silicon is $\sim$60K. Warm up is a slow procedure so the risk of entering leakage was mitigated for all these measurements.
}

\bibliography{bibli}

\begin{thebibliography}{48}%
\makeatletter
\providecommand \@ifxundefined [1]{%
 \@ifx{#1\undefined}
}%
\providecommand \@ifnum [1]{%
 \ifnum #1\expandafter \@firstoftwo
 \else \expandafter \@secondoftwo
 \fi
}%
\providecommand \@ifx [1]{%
 \ifx #1\expandafter \@firstoftwo
 \else \expandafter \@secondoftwo
 \fi
}%
\providecommand \natexlab [1]{#1}%
\providecommand \enquote  [1]{``#1''}%
\providecommand \bibnamefont  [1]{#1}%
\providecommand \bibfnamefont [1]{#1}%
\providecommand \citenamefont [1]{#1}%
\providecommand \href@noop [0]{\@secondoftwo}%
\providecommand \href [0]{\begingroup \@sanitize@url \@href}%
\providecommand \@href[1]{\@@startlink{#1}\@@href}%
\providecommand \@@href[1]{\endgroup#1\@@endlink}%
\providecommand \@sanitize@url [0]{\catcode `\\12\catcode `\$12\catcode `\&12\catcode `\#12\catcode `\^12\catcode `\_12\catcode `\%12\relax}%
\providecommand \@@startlink[1]{}%
\providecommand \@@endlink[0]{}%
\providecommand \url  [0]{\begingroup\@sanitize@url \@url }%
\providecommand \@url [1]{\endgroup\@href {#1}{\urlprefix }}%
\providecommand \urlprefix  [0]{URL }%
\providecommand \Eprint [0]{\href }%
\providecommand \doibase [0]{https://doi.org/}%
\providecommand \selectlanguage [0]{\@gobble}%
\providecommand \bibinfo  [0]{\@secondoftwo}%
\providecommand \bibfield  [0]{\@secondoftwo}%
\providecommand \translation [1]{[#1]}%
\providecommand \BibitemOpen [0]{}%
\providecommand \bibitemStop [0]{}%
\providecommand \bibitemNoStop [0]{.\EOS\space}%
\providecommand \EOS [0]{\spacefactor3000\relax}%
\providecommand \BibitemShut  [1]{\csname bibitem#1\endcsname}%
\let\auto@bib@innerbib\@empty
\bibitem [{\citenamefont {Fowler}\ \emph {et~al.}(2012)\citenamefont {Fowler}, \citenamefont {Mariantoni}, \citenamefont {Martinis},\ and\ \citenamefont {Cleland}}]{Fowler2012}%
  \BibitemOpen
  \bibfield  {author} {\bibinfo {author} {\bibfnamefont {A.~G.}\ \bibnamefont {Fowler}}, \bibinfo {author} {\bibfnamefont {M.}~\bibnamefont {Mariantoni}}, \bibinfo {author} {\bibfnamefont {J.~M.}\ \bibnamefont {Martinis}},\ and\ \bibinfo {author} {\bibfnamefont {A.~N.}\ \bibnamefont {Cleland}},\ }\bibfield  {title} {\bibinfo {title} {Surface codes: Towards practical large-scale quantum computation},\ }\href {https://doi.org/10.1103/PhysRevA.86.032324} {\bibfield  {journal} {\bibinfo  {journal} {Physical Review A}\ }\textbf {\bibinfo {volume} {86}},\ \bibinfo {pages} {032324} (\bibinfo {year} {2012})}\BibitemShut {NoStop}%
\bibitem [{\citenamefont {Fowler}\ and\ \citenamefont {Martinis}(2014)}]{Fowler2014}%
  \BibitemOpen
  \bibfield  {author} {\bibinfo {author} {\bibfnamefont {A.~G.}\ \bibnamefont {Fowler}}\ and\ \bibinfo {author} {\bibfnamefont {J.~M.}\ \bibnamefont {Martinis}},\ }\bibfield  {title} {\bibinfo {title} {Quantifying the effects of local many-qubit errors and nonlocal two-qubit errors on the surface code},\ }\href@noop {} {\bibfield  {journal} {\bibinfo  {journal} {Physical Review A}\ }\textbf {\bibinfo {volume} {89}},\ \bibinfo {pages} {032316} (\bibinfo {year} {2014})}\BibitemShut {NoStop}%
\bibitem [{\citenamefont {Clemens}\ \emph {et~al.}(2004)\citenamefont {Clemens}, \citenamefont {Siddiqui},\ and\ \citenamefont {Gea-Banacloche}}]{Clemens2004}%
  \BibitemOpen
  \bibfield  {author} {\bibinfo {author} {\bibfnamefont {J.~P.}\ \bibnamefont {Clemens}}, \bibinfo {author} {\bibfnamefont {S.}~\bibnamefont {Siddiqui}},\ and\ \bibinfo {author} {\bibfnamefont {J.}~\bibnamefont {Gea-Banacloche}},\ }\bibfield  {title} {\bibinfo {title} {Quantum error correction against correlated noise},\ }\href@noop {} {\bibfield  {journal} {\bibinfo  {journal} {Physical Review A}\ }\textbf {\bibinfo {volume} {69}},\ \bibinfo {pages} {062313} (\bibinfo {year} {2004})}\BibitemShut {NoStop}%
\bibitem [{\citenamefont {Aharonov}\ \emph {et~al.}(2006)\citenamefont {Aharonov}, \citenamefont {Kitaev},\ and\ \citenamefont {Preskill}}]{Aharonov2006}%
  \BibitemOpen
  \bibfield  {author} {\bibinfo {author} {\bibfnamefont {D.}~\bibnamefont {Aharonov}}, \bibinfo {author} {\bibfnamefont {A.}~\bibnamefont {Kitaev}},\ and\ \bibinfo {author} {\bibfnamefont {J.}~\bibnamefont {Preskill}},\ }\bibfield  {title} {\bibinfo {title} {Fault-tolerant quantum computation with long-range correlated noise},\ }\href@noop {} {\bibfield  {journal} {\bibinfo  {journal} {Physical Review Letters}\ }\textbf {\bibinfo {volume} {96}},\ \bibinfo {pages} {050504} (\bibinfo {year} {2006})}\BibitemShut {NoStop}%
\bibitem [{\citenamefont {Wilen}\ \emph {et~al.}(2021)\citenamefont {Wilen}, \citenamefont {Abdullah}, \citenamefont {Kurinsky}, \citenamefont {Stanford}, \citenamefont {Cardani}, \citenamefont {d’Imperio}, \citenamefont {Tomei}, \citenamefont {Faoro}, \citenamefont {Ioffe}, \citenamefont {Liu} \emph {et~al.}}]{wilen2021}%
  \BibitemOpen
  \bibfield  {author} {\bibinfo {author} {\bibfnamefont {C.~D.}\ \bibnamefont {Wilen}}, \bibinfo {author} {\bibfnamefont {S.}~\bibnamefont {Abdullah}}, \bibinfo {author} {\bibfnamefont {N.}~\bibnamefont {Kurinsky}}, \bibinfo {author} {\bibfnamefont {C.}~\bibnamefont {Stanford}}, \bibinfo {author} {\bibfnamefont {L.}~\bibnamefont {Cardani}}, \bibinfo {author} {\bibfnamefont {G.}~\bibnamefont {d’Imperio}}, \bibinfo {author} {\bibfnamefont {C.}~\bibnamefont {Tomei}}, \bibinfo {author} {\bibfnamefont {L.}~\bibnamefont {Faoro}}, \bibinfo {author} {\bibfnamefont {L.}~\bibnamefont {Ioffe}}, \bibinfo {author} {\bibfnamefont {C.}~\bibnamefont {Liu}}, \emph {et~al.},\ }\bibfield  {title} {\bibinfo {title} {Correlated charge noise and relaxation errors in superconducting qubits},\ }\href@noop {} {\bibfield  {journal} {\bibinfo  {journal} {Nature}\ }\textbf {\bibinfo {volume} {594}},\ \bibinfo {pages} {369} (\bibinfo {year} {2021})}\BibitemShut {NoStop}%
\bibitem [{\citenamefont {Martinis}(2021)}]{Martinis2021}%
  \BibitemOpen
  \bibfield  {author} {\bibinfo {author} {\bibfnamefont {J.~M.}\ \bibnamefont {Martinis}},\ }\bibfield  {title} {\bibinfo {title} {Saving superconducting quantum processors from decay and correlated errors generated by gamma and cosmic rays},\ }\href@noop {} {\bibfield  {journal} {\bibinfo  {journal} {npj Quantum Information}\ }\textbf {\bibinfo {volume} {7}},\ \bibinfo {pages} {90} (\bibinfo {year} {2021})}\BibitemShut {NoStop}%
\bibitem [{\citenamefont {McEwen}\ \emph {et~al.}(2022)\citenamefont {McEwen}, \citenamefont {Faoro}, \citenamefont {Arya}, \citenamefont {Dunsworth}, \citenamefont {Huang}, \citenamefont {Kim}, \citenamefont {Burkett}, \citenamefont {Fowler}, \citenamefont {Arute}, \citenamefont {Bardin} \emph {et~al.}}]{Mcewen2022}%
  \BibitemOpen
  \bibfield  {author} {\bibinfo {author} {\bibfnamefont {M.}~\bibnamefont {McEwen}}, \bibinfo {author} {\bibfnamefont {L.}~\bibnamefont {Faoro}}, \bibinfo {author} {\bibfnamefont {K.}~\bibnamefont {Arya}}, \bibinfo {author} {\bibfnamefont {A.}~\bibnamefont {Dunsworth}}, \bibinfo {author} {\bibfnamefont {T.}~\bibnamefont {Huang}}, \bibinfo {author} {\bibfnamefont {S.}~\bibnamefont {Kim}}, \bibinfo {author} {\bibfnamefont {B.}~\bibnamefont {Burkett}}, \bibinfo {author} {\bibfnamefont {A.}~\bibnamefont {Fowler}}, \bibinfo {author} {\bibfnamefont {F.}~\bibnamefont {Arute}}, \bibinfo {author} {\bibfnamefont {J.~C.}\ \bibnamefont {Bardin}}, \emph {et~al.},\ }\bibfield  {title} {\bibinfo {title} {Resolving catastrophic error bursts from cosmic rays in large arrays of superconducting qubits},\ }\href@noop {} {\bibfield  {journal} {\bibinfo  {journal} {Nature Physics}\ }\textbf {\bibinfo {volume} {18}},\ \bibinfo {pages} {107} (\bibinfo {year} {2022})}\BibitemShut {NoStop}%
\bibitem [{\citenamefont {{Google Quantum AI}}(2021)}]{google2021}%
  \BibitemOpen
  \bibfield  {author} {\bibinfo {author} {\bibnamefont {{Google Quantum AI}}},\ }\bibfield  {title} {\bibinfo {title} {Exponential suppression of bit or phase errors with cyclic error correction},\ }\href@noop {} {\bibfield  {journal} {\bibinfo  {journal} {Nature}\ }\textbf {\bibinfo {volume} {595}},\ \bibinfo {pages} {383} (\bibinfo {year} {2021})}\BibitemShut {NoStop}%
\bibitem [{\citenamefont {Patel}\ \emph {et~al.}(2017)\citenamefont {Patel}, \citenamefont {Pechenezhskiy}, \citenamefont {Plourde}, \citenamefont {Vavilov},\ and\ \citenamefont {McDermott}}]{Patel2017}%
  \BibitemOpen
  \bibfield  {author} {\bibinfo {author} {\bibfnamefont {U.}~\bibnamefont {Patel}}, \bibinfo {author} {\bibfnamefont {I.~V.}\ \bibnamefont {Pechenezhskiy}}, \bibinfo {author} {\bibfnamefont {B.}~\bibnamefont {Plourde}}, \bibinfo {author} {\bibfnamefont {M.}~\bibnamefont {Vavilov}},\ and\ \bibinfo {author} {\bibfnamefont {R.}~\bibnamefont {McDermott}},\ }\bibfield  {title} {\bibinfo {title} {Phonon-mediated quasiparticle poisoning of superconducting microwave resonators},\ }\href@noop {} {\bibfield  {journal} {\bibinfo  {journal} {Physical Review B}\ }\textbf {\bibinfo {volume} {96}},\ \bibinfo {pages} {220501} (\bibinfo {year} {2017})}\BibitemShut {NoStop}%
\bibitem [{\citenamefont {Liu}\ \emph {et~al.}(2022)\citenamefont {Liu}, \citenamefont {Harrison}, \citenamefont {Patel}, \citenamefont {Wilen}, \citenamefont {Rafferty}, \citenamefont {Shearrow}, \citenamefont {Ballard}, \citenamefont {Iaia}, \citenamefont {Ku}, \citenamefont {Plourde} \emph {et~al.}}]{Liu2022}%
  \BibitemOpen
  \bibfield  {author} {\bibinfo {author} {\bibfnamefont {C.-H.}\ \bibnamefont {Liu}}, \bibinfo {author} {\bibfnamefont {D.~C.}\ \bibnamefont {Harrison}}, \bibinfo {author} {\bibfnamefont {S.}~\bibnamefont {Patel}}, \bibinfo {author} {\bibfnamefont {C.~D.}\ \bibnamefont {Wilen}}, \bibinfo {author} {\bibfnamefont {O.}~\bibnamefont {Rafferty}}, \bibinfo {author} {\bibfnamefont {A.}~\bibnamefont {Shearrow}}, \bibinfo {author} {\bibfnamefont {A.}~\bibnamefont {Ballard}}, \bibinfo {author} {\bibfnamefont {V.}~\bibnamefont {Iaia}}, \bibinfo {author} {\bibfnamefont {J.}~\bibnamefont {Ku}}, \bibinfo {author} {\bibfnamefont {B.~L.}\ \bibnamefont {Plourde}}, \emph {et~al.},\ }\bibfield  {title} {\bibinfo {title} {Quasiparticle poisoning of superconducting qubits from resonant absorption of pair-breaking photons},\ }\href@noop {} {\bibfield  {journal} {\bibinfo  {journal} {arXiv preprint arXiv:2203.06577}\ } (\bibinfo {year} {2022})}\BibitemShut {NoStop}%
\bibitem [{\citenamefont {M{\k{a}}dzik}\ \emph {et~al.}(2022)\citenamefont {M{\k{a}}dzik}, \citenamefont {Asaad}, \citenamefont {Youssry}, \citenamefont {Joecker}, \citenamefont {Rudinger}, \citenamefont {Nielsen}, \citenamefont {Young}, \citenamefont {Proctor}, \citenamefont {Baczewski}, \citenamefont {Laucht} \emph {et~al.}}]{mkadzik2022}%
  \BibitemOpen
  \bibfield  {author} {\bibinfo {author} {\bibfnamefont {M.~T.}\ \bibnamefont {M{\k{a}}dzik}}, \bibinfo {author} {\bibfnamefont {S.}~\bibnamefont {Asaad}}, \bibinfo {author} {\bibfnamefont {A.}~\bibnamefont {Youssry}}, \bibinfo {author} {\bibfnamefont {B.}~\bibnamefont {Joecker}}, \bibinfo {author} {\bibfnamefont {K.~M.}\ \bibnamefont {Rudinger}}, \bibinfo {author} {\bibfnamefont {E.}~\bibnamefont {Nielsen}}, \bibinfo {author} {\bibfnamefont {K.~C.}\ \bibnamefont {Young}}, \bibinfo {author} {\bibfnamefont {T.~J.}\ \bibnamefont {Proctor}}, \bibinfo {author} {\bibfnamefont {A.~D.}\ \bibnamefont {Baczewski}}, \bibinfo {author} {\bibfnamefont {A.}~\bibnamefont {Laucht}}, \emph {et~al.},\ }\bibfield  {title} {\bibinfo {title} {Precision tomography of a three-qubit donor quantum processor in silicon},\ }\href@noop {} {\bibfield  {journal} {\bibinfo  {journal} {Nature}\ }\textbf {\bibinfo {volume} {601}},\ \bibinfo {pages} {348} (\bibinfo {year} {2022})}\BibitemShut {NoStop}%
\bibitem [{\citenamefont {Noiri}\ \emph {et~al.}(2022)\citenamefont {Noiri}, \citenamefont {Takeda}, \citenamefont {Nakajima}, \citenamefont {Kobayashi}, \citenamefont {Sammak}, \citenamefont {Scappucci},\ and\ \citenamefont {Tarucha}}]{noiri2022}%
  \BibitemOpen
  \bibfield  {author} {\bibinfo {author} {\bibfnamefont {A.}~\bibnamefont {Noiri}}, \bibinfo {author} {\bibfnamefont {K.}~\bibnamefont {Takeda}}, \bibinfo {author} {\bibfnamefont {T.}~\bibnamefont {Nakajima}}, \bibinfo {author} {\bibfnamefont {T.}~\bibnamefont {Kobayashi}}, \bibinfo {author} {\bibfnamefont {A.}~\bibnamefont {Sammak}}, \bibinfo {author} {\bibfnamefont {G.}~\bibnamefont {Scappucci}},\ and\ \bibinfo {author} {\bibfnamefont {S.}~\bibnamefont {Tarucha}},\ }\bibfield  {title} {\bibinfo {title} {Fast universal quantum gate above the fault-tolerance threshold in silicon},\ }\href@noop {} {\bibfield  {journal} {\bibinfo  {journal} {Nature}\ }\textbf {\bibinfo {volume} {601}},\ \bibinfo {pages} {338} (\bibinfo {year} {2022})}\BibitemShut {NoStop}%
\bibitem [{\citenamefont {Xue}\ \emph {et~al.}(2022)\citenamefont {Xue}, \citenamefont {Russ}, \citenamefont {Samkharadze}, \citenamefont {Undseth}, \citenamefont {Sammak}, \citenamefont {Scappucci},\ and\ \citenamefont {Vandersypen}}]{xue2022}%
  \BibitemOpen
  \bibfield  {author} {\bibinfo {author} {\bibfnamefont {X.}~\bibnamefont {Xue}}, \bibinfo {author} {\bibfnamefont {M.}~\bibnamefont {Russ}}, \bibinfo {author} {\bibfnamefont {N.}~\bibnamefont {Samkharadze}}, \bibinfo {author} {\bibfnamefont {B.}~\bibnamefont {Undseth}}, \bibinfo {author} {\bibfnamefont {A.}~\bibnamefont {Sammak}}, \bibinfo {author} {\bibfnamefont {G.}~\bibnamefont {Scappucci}},\ and\ \bibinfo {author} {\bibfnamefont {L.~M.}\ \bibnamefont {Vandersypen}},\ }\bibfield  {title} {\bibinfo {title} {Quantum logic with spin qubits crossing the surface code threshold},\ }\href@noop {} {\bibfield  {journal} {\bibinfo  {journal} {Nature}\ }\textbf {\bibinfo {volume} {601}},\ \bibinfo {pages} {343} (\bibinfo {year} {2022})}\BibitemShut {NoStop}%
\bibitem [{\citenamefont {Connors}\ \emph {et~al.}(2019)\citenamefont {Connors}, \citenamefont {Nelson}, \citenamefont {Qiao}, \citenamefont {Edge},\ and\ \citenamefont {Nichol}}]{Connors2019}%
  \BibitemOpen
  \bibfield  {author} {\bibinfo {author} {\bibfnamefont {E.~J.}\ \bibnamefont {Connors}}, \bibinfo {author} {\bibfnamefont {J.}~\bibnamefont {Nelson}}, \bibinfo {author} {\bibfnamefont {H.}~\bibnamefont {Qiao}}, \bibinfo {author} {\bibfnamefont {L.~F.}\ \bibnamefont {Edge}},\ and\ \bibinfo {author} {\bibfnamefont {J.~M.}\ \bibnamefont {Nichol}},\ }\bibfield  {title} {\bibinfo {title} {Low-frequency charge noise in si/sige quantum dots},\ }\href@noop {} {\bibfield  {journal} {\bibinfo  {journal} {Physical Review B}\ }\textbf {\bibinfo {volume} {100}},\ \bibinfo {pages} {165305} (\bibinfo {year} {2019})}\BibitemShut {NoStop}%
\bibitem [{\citenamefont {Freeman}\ \emph {et~al.}(2016)\citenamefont {Freeman}, \citenamefont {Schoenfield},\ and\ \citenamefont {Jiang}}]{Freeman2016}%
  \BibitemOpen
  \bibfield  {author} {\bibinfo {author} {\bibfnamefont {B.~M.}\ \bibnamefont {Freeman}}, \bibinfo {author} {\bibfnamefont {J.~S.}\ \bibnamefont {Schoenfield}},\ and\ \bibinfo {author} {\bibfnamefont {H.}~\bibnamefont {Jiang}},\ }\bibfield  {title} {\bibinfo {title} {Comparison of low frequency charge noise in identically patterned si/sio2 and si/sige quantum dots},\ }\href@noop {} {\bibfield  {journal} {\bibinfo  {journal} {Applied Physics Letters}\ }\textbf {\bibinfo {volume} {108}},\ \bibinfo {pages} {253108} (\bibinfo {year} {2016})}\BibitemShut {NoStop}%
\bibitem [{\citenamefont {Zimmerman}\ \emph {et~al.}(2014)\citenamefont {Zimmerman}, \citenamefont {Yang}, \citenamefont {Lai}, \citenamefont {Lim},\ and\ \citenamefont {Dzurak}}]{zimmerman2014}%
  \BibitemOpen
  \bibfield  {author} {\bibinfo {author} {\bibfnamefont {N.~M.}\ \bibnamefont {Zimmerman}}, \bibinfo {author} {\bibfnamefont {C.-H.}\ \bibnamefont {Yang}}, \bibinfo {author} {\bibfnamefont {N.~S.}\ \bibnamefont {Lai}}, \bibinfo {author} {\bibfnamefont {W.~H.}\ \bibnamefont {Lim}},\ and\ \bibinfo {author} {\bibfnamefont {A.~S.}\ \bibnamefont {Dzurak}},\ }\bibfield  {title} {\bibinfo {title} {Charge offset stability in si single electron devices with al gates},\ }\href@noop {} {\bibfield  {journal} {\bibinfo  {journal} {Nanotechnology}\ }\textbf {\bibinfo {volume} {25}},\ \bibinfo {pages} {405201} (\bibinfo {year} {2014})}\BibitemShut {NoStop}%
\bibitem [{\citenamefont {Rudolph}\ \emph {et~al.}(2019)\citenamefont {Rudolph}, \citenamefont {Sarabi}, \citenamefont {Murray}, \citenamefont {Carroll},\ and\ \citenamefont {Zimmerman}}]{Rudolph2019}%
  \BibitemOpen
  \bibfield  {author} {\bibinfo {author} {\bibfnamefont {M.}~\bibnamefont {Rudolph}}, \bibinfo {author} {\bibfnamefont {B.}~\bibnamefont {Sarabi}}, \bibinfo {author} {\bibfnamefont {R.}~\bibnamefont {Murray}}, \bibinfo {author} {\bibfnamefont {M.}~\bibnamefont {Carroll}},\ and\ \bibinfo {author} {\bibfnamefont {N.~M.}\ \bibnamefont {Zimmerman}},\ }\bibfield  {title} {\bibinfo {title} {Long-term drift of si-mos quantum dots with intentional donor implants},\ }\href@noop {} {\bibfield  {journal} {\bibinfo  {journal} {Scientific Reports}\ }\textbf {\bibinfo {volume} {9}},\ \bibinfo {pages} {7656} (\bibinfo {year} {2019})}\BibitemShut {NoStop}%
\bibitem [{\citenamefont {Takeda}\ \emph {et~al.}(2013)\citenamefont {Takeda}, \citenamefont {Obata}, \citenamefont {Fukuoka}, \citenamefont {Akhtar}, \citenamefont {Kamioka}, \citenamefont {Kodera}, \citenamefont {Oda},\ and\ \citenamefont {Tarucha}}]{takeda2013}%
  \BibitemOpen
  \bibfield  {author} {\bibinfo {author} {\bibfnamefont {K.}~\bibnamefont {Takeda}}, \bibinfo {author} {\bibfnamefont {T.}~\bibnamefont {Obata}}, \bibinfo {author} {\bibfnamefont {Y.}~\bibnamefont {Fukuoka}}, \bibinfo {author} {\bibfnamefont {W.}~\bibnamefont {Akhtar}}, \bibinfo {author} {\bibfnamefont {J.}~\bibnamefont {Kamioka}}, \bibinfo {author} {\bibfnamefont {T.}~\bibnamefont {Kodera}}, \bibinfo {author} {\bibfnamefont {S.}~\bibnamefont {Oda}},\ and\ \bibinfo {author} {\bibfnamefont {S.}~\bibnamefont {Tarucha}},\ }\bibfield  {title} {\bibinfo {title} {Characterization and suppression of low-frequency noise in si/sige quantum point contacts and quantum dots},\ }\href@noop {} {\bibfield  {journal} {\bibinfo  {journal} {Applied Physics Letters}\ }\textbf {\bibinfo {volume} {102}},\ \bibinfo {pages} {123113} (\bibinfo {year} {2013})}\BibitemShut {NoStop}%
\bibitem [{\citenamefont {Connors}\ \emph {et~al.}(2022)\citenamefont {Connors}, \citenamefont {Nelson}, \citenamefont {Edge},\ and\ \citenamefont {Nichol}}]{Connors2022}%
  \BibitemOpen
  \bibfield  {author} {\bibinfo {author} {\bibfnamefont {E.~J.}\ \bibnamefont {Connors}}, \bibinfo {author} {\bibfnamefont {J.}~\bibnamefont {Nelson}}, \bibinfo {author} {\bibfnamefont {L.~F.}\ \bibnamefont {Edge}},\ and\ \bibinfo {author} {\bibfnamefont {J.~M.}\ \bibnamefont {Nichol}},\ }\bibfield  {title} {\bibinfo {title} {Charge-noise spectroscopy of {Si/SiGe} quantum dots via dynamically-decoupled exchange oscillations},\ }\href@noop {} {\bibfield  {journal} {\bibinfo  {journal} {Nature Communications}\ }\textbf {\bibinfo {volume} {13}},\ \bibinfo {pages} {940} (\bibinfo {year} {2022})}\BibitemShut {NoStop}%
\bibitem [{\citenamefont {Kranz}\ \emph {et~al.}(2020)\citenamefont {Kranz}, \citenamefont {Gorman}, \citenamefont {Thorgrimsson}, \citenamefont {He}, \citenamefont {Keith}, \citenamefont {Keizer},\ and\ \citenamefont {Simmons}}]{Kranz2020}%
  \BibitemOpen
  \bibfield  {author} {\bibinfo {author} {\bibfnamefont {L.}~\bibnamefont {Kranz}}, \bibinfo {author} {\bibfnamefont {S.~K.}\ \bibnamefont {Gorman}}, \bibinfo {author} {\bibfnamefont {B.}~\bibnamefont {Thorgrimsson}}, \bibinfo {author} {\bibfnamefont {Y.}~\bibnamefont {He}}, \bibinfo {author} {\bibfnamefont {D.}~\bibnamefont {Keith}}, \bibinfo {author} {\bibfnamefont {J.~G.}\ \bibnamefont {Keizer}},\ and\ \bibinfo {author} {\bibfnamefont {M.~Y.}\ \bibnamefont {Simmons}},\ }\bibfield  {title} {\bibinfo {title} {Exploiting a single-crystal environment to minimize the charge noise on qubits in silicon},\ }\href {https://doi.org/https://doi.org/10.1002/adma.202003361} {\bibfield  {journal} {\bibinfo  {journal} {Advanced Materials}\ }\textbf {\bibinfo {volume} {32}},\ \bibinfo {pages} {2003361} (\bibinfo {year} {2020})},\ \Eprint {https://arxiv.org/abs/https://onlinelibrary.wiley.com/doi/pdf/10.1002/adma.202003361} {https://onlinelibrary.wiley.com/doi/pdf/10.1002/adma.202003361} \BibitemShut {NoStop}%
\bibitem [{\citenamefont {Boter}\ \emph {et~al.}(2020)\citenamefont {Boter}, \citenamefont {Xue}, \citenamefont {Kr\"ahenmann}, \citenamefont {Watson}, \citenamefont {Premakumar}, \citenamefont {Ward}, \citenamefont {Savage}, \citenamefont {Lagally}, \citenamefont {Friesen}, \citenamefont {Coppersmith}, \citenamefont {Eriksson}, \citenamefont {Joynt},\ and\ \citenamefont {Vandersypen}}]{Boter2020}%
  \BibitemOpen
  \bibfield  {author} {\bibinfo {author} {\bibfnamefont {J.~M.}\ \bibnamefont {Boter}}, \bibinfo {author} {\bibfnamefont {X.}~\bibnamefont {Xue}}, \bibinfo {author} {\bibfnamefont {T.}~\bibnamefont {Kr\"ahenmann}}, \bibinfo {author} {\bibfnamefont {T.~F.}\ \bibnamefont {Watson}}, \bibinfo {author} {\bibfnamefont {V.~N.}\ \bibnamefont {Premakumar}}, \bibinfo {author} {\bibfnamefont {D.~R.}\ \bibnamefont {Ward}}, \bibinfo {author} {\bibfnamefont {D.~E.}\ \bibnamefont {Savage}}, \bibinfo {author} {\bibfnamefont {M.~G.}\ \bibnamefont {Lagally}}, \bibinfo {author} {\bibfnamefont {M.}~\bibnamefont {Friesen}}, \bibinfo {author} {\bibfnamefont {S.~N.}\ \bibnamefont {Coppersmith}}, \bibinfo {author} {\bibfnamefont {M.~A.}\ \bibnamefont {Eriksson}}, \bibinfo {author} {\bibfnamefont {R.}~\bibnamefont {Joynt}},\ and\ \bibinfo {author} {\bibfnamefont {L.~M.~K.}\ \bibnamefont {Vandersypen}},\ }\bibfield  {title} {\bibinfo {title} {Spatial noise correlations in a si/sige two-qubit device from bell state coherences},\ }\href
  {https://doi.org/10.1103/PhysRevB.101.235133} {\bibfield  {journal} {\bibinfo  {journal} {Physical Review B}\ }\textbf {\bibinfo {volume} {101}},\ \bibinfo {pages} {235133} (\bibinfo {year} {2020})}\BibitemShut {NoStop}%
\bibitem [{\citenamefont {Yoneda}\ \emph {et~al.}(2022)\citenamefont {Yoneda}, \citenamefont {Rojas-Arias}, \citenamefont {Stano}, \citenamefont {Takeda}, \citenamefont {Noiri}, \citenamefont {Nakajima}, \citenamefont {Loss},\ and\ \citenamefont {Tarucha}}]{Yoneda2022}%
  \BibitemOpen
  \bibfield  {author} {\bibinfo {author} {\bibfnamefont {J.}~\bibnamefont {Yoneda}}, \bibinfo {author} {\bibfnamefont {J.}~\bibnamefont {Rojas-Arias}}, \bibinfo {author} {\bibfnamefont {P.}~\bibnamefont {Stano}}, \bibinfo {author} {\bibfnamefont {K.}~\bibnamefont {Takeda}}, \bibinfo {author} {\bibfnamefont {A.}~\bibnamefont {Noiri}}, \bibinfo {author} {\bibfnamefont {T.}~\bibnamefont {Nakajima}}, \bibinfo {author} {\bibfnamefont {D.}~\bibnamefont {Loss}},\ and\ \bibinfo {author} {\bibfnamefont {S.}~\bibnamefont {Tarucha}},\ }\bibfield  {title} {\bibinfo {title} {Noise-correlation spectrum for a pair of spin qubits in silicon},\ }\href@noop {} {\bibfield  {journal} {\bibinfo  {journal} {arXiv preprint arXiv:2208.14150}\ } (\bibinfo {year} {2022})}\BibitemShut {NoStop}%
\bibitem [{\citenamefont {Rojas-Arias}\ \emph {et~al.}(2023)\citenamefont {Rojas-Arias}, \citenamefont {Noiri}, \citenamefont {Stano}, \citenamefont {Nakajima}, \citenamefont {Yoneda}, \citenamefont {Takeda}, \citenamefont {Kobayashi}, \citenamefont {Sammak}, \citenamefont {Scappucci}, \citenamefont {Loss} \emph {et~al.}}]{rojasarias2023}%
  \BibitemOpen
  \bibfield  {author} {\bibinfo {author} {\bibfnamefont {J.~S.}\ \bibnamefont {Rojas-Arias}}, \bibinfo {author} {\bibfnamefont {A.}~\bibnamefont {Noiri}}, \bibinfo {author} {\bibfnamefont {P.}~\bibnamefont {Stano}}, \bibinfo {author} {\bibfnamefont {T.}~\bibnamefont {Nakajima}}, \bibinfo {author} {\bibfnamefont {J.}~\bibnamefont {Yoneda}}, \bibinfo {author} {\bibfnamefont {K.}~\bibnamefont {Takeda}}, \bibinfo {author} {\bibfnamefont {T.}~\bibnamefont {Kobayashi}}, \bibinfo {author} {\bibfnamefont {A.}~\bibnamefont {Sammak}}, \bibinfo {author} {\bibfnamefont {G.}~\bibnamefont {Scappucci}}, \bibinfo {author} {\bibfnamefont {D.}~\bibnamefont {Loss}}, \emph {et~al.},\ }\bibfield  {title} {\bibinfo {title} {Spatial noise correlations beyond nearest-neighbor in $^{28}$si/sige spin qubits},\ }\href@noop {} {\bibfield  {journal} {\bibinfo  {journal} {arXiv preprint arXiv:2302.11717}\ } (\bibinfo {year} {2023})}\BibitemShut {NoStop}%
\bibitem [{\citenamefont {Pakkiam}\ \emph {et~al.}(2018)\citenamefont {Pakkiam}, \citenamefont {House}, \citenamefont {Koch},\ and\ \citenamefont {Simmons}}]{Pakkiam2018}%
  \BibitemOpen
  \bibfield  {author} {\bibinfo {author} {\bibfnamefont {P.}~\bibnamefont {Pakkiam}}, \bibinfo {author} {\bibfnamefont {M.~G.}\ \bibnamefont {House}}, \bibinfo {author} {\bibfnamefont {M.}~\bibnamefont {Koch}},\ and\ \bibinfo {author} {\bibfnamefont {M.~Y.}\ \bibnamefont {Simmons}},\ }\bibfield  {title} {\bibinfo {title} {Characterization of a scalable donor-based singlet-triplet qubit architecture in silicon},\ }\href {https://doi.org/10.1021/acs.nanolett.8b00006} {\bibfield  {journal} {\bibinfo  {journal} {Nano Letters}\ }\textbf {\bibinfo {volume} {18}},\ \bibinfo {pages} {4081} (\bibinfo {year} {2018})}\BibitemShut {NoStop}%
\bibitem [{\citenamefont {Shamim}\ \emph {et~al.}(2016)\citenamefont {Shamim}, \citenamefont {Weber}, \citenamefont {Thompson}, \citenamefont {Simmons},\ and\ \citenamefont {Ghosh}}]{Shamim2016}%
  \BibitemOpen
  \bibfield  {author} {\bibinfo {author} {\bibfnamefont {S.}~\bibnamefont {Shamim}}, \bibinfo {author} {\bibfnamefont {B.}~\bibnamefont {Weber}}, \bibinfo {author} {\bibfnamefont {D.~W.}\ \bibnamefont {Thompson}}, \bibinfo {author} {\bibfnamefont {M.~Y.}\ \bibnamefont {Simmons}},\ and\ \bibinfo {author} {\bibfnamefont {A.}~\bibnamefont {Ghosh}},\ }\bibfield  {title} {\bibinfo {title} {Ultralow-noise atomic-scale structures for quantum circuitry in silicon},\ }\href {https://doi.org/10.1021/acs.nanolett.6b02513} {\bibfield  {journal} {\bibinfo  {journal} {Nano Letters}\ }\textbf {\bibinfo {volume} {16}},\ \bibinfo {pages} {5779} (\bibinfo {year} {2016})}\BibitemShut {NoStop}%
\bibitem [{\citenamefont {Yoneda}\ \emph {et~al.}(2018)\citenamefont {Yoneda}, \citenamefont {Takeda}, \citenamefont {Otsuka}, \citenamefont {Nakajima}, \citenamefont {Delbecq}, \citenamefont {Allison}, \citenamefont {Honda}, \citenamefont {Kodera}, \citenamefont {Oda}, \citenamefont {Hoshi} \emph {et~al.}}]{Yoneda2018}%
  \BibitemOpen
  \bibfield  {author} {\bibinfo {author} {\bibfnamefont {J.}~\bibnamefont {Yoneda}}, \bibinfo {author} {\bibfnamefont {K.}~\bibnamefont {Takeda}}, \bibinfo {author} {\bibfnamefont {T.}~\bibnamefont {Otsuka}}, \bibinfo {author} {\bibfnamefont {T.}~\bibnamefont {Nakajima}}, \bibinfo {author} {\bibfnamefont {M.~R.}\ \bibnamefont {Delbecq}}, \bibinfo {author} {\bibfnamefont {G.}~\bibnamefont {Allison}}, \bibinfo {author} {\bibfnamefont {T.}~\bibnamefont {Honda}}, \bibinfo {author} {\bibfnamefont {T.}~\bibnamefont {Kodera}}, \bibinfo {author} {\bibfnamefont {S.}~\bibnamefont {Oda}}, \bibinfo {author} {\bibfnamefont {Y.}~\bibnamefont {Hoshi}}, \emph {et~al.},\ }\bibfield  {title} {\bibinfo {title} {A quantum-dot spin qubit with coherence limited by charge noise and fidelity higher than 99.9\%},\ }\href@noop {} {\bibfield  {journal} {\bibinfo  {journal} {Nature Nanotechnology}\ }\textbf {\bibinfo {volume} {13}},\ \bibinfo {pages} {102} (\bibinfo {year} {2018})}\BibitemShut {NoStop}%
\bibitem [{\citenamefont {He}\ \emph {et~al.}(2019)\citenamefont {He}, \citenamefont {Gorman}, \citenamefont {Keith}, \citenamefont {Kranz}, \citenamefont {Keizer},\ and\ \citenamefont {Simmons}}]{He2019}%
  \BibitemOpen
  \bibfield  {author} {\bibinfo {author} {\bibfnamefont {Y.}~\bibnamefont {He}}, \bibinfo {author} {\bibfnamefont {S.~K.}\ \bibnamefont {Gorman}}, \bibinfo {author} {\bibfnamefont {D.}~\bibnamefont {Keith}}, \bibinfo {author} {\bibfnamefont {L.}~\bibnamefont {Kranz}}, \bibinfo {author} {\bibfnamefont {J.~G.}\ \bibnamefont {Keizer}},\ and\ \bibinfo {author} {\bibfnamefont {M.~Y.}\ \bibnamefont {Simmons}},\ }\bibfield  {title} {\bibinfo {title} {A two-qubit gate between phosphorus donor electrons in silicon},\ }\href {https://doi.org/10.1038/s41586-019-1381-2} {\bibfield  {journal} {\bibinfo  {journal} {Nature}\ }\textbf {\bibinfo {volume} {571}},\ \bibinfo {pages} {371} (\bibinfo {year} {2019})}\BibitemShut {NoStop}%
\bibitem [{\citenamefont {Koch}\ \emph {et~al.}(2019)\citenamefont {Koch}, \citenamefont {Keizer}, \citenamefont {Pakkiam}, \citenamefont {Keith}, \citenamefont {House}, \citenamefont {Peretz},\ and\ \citenamefont {Simmons}}]{Koch2019}%
  \BibitemOpen
  \bibfield  {author} {\bibinfo {author} {\bibfnamefont {M.}~\bibnamefont {Koch}}, \bibinfo {author} {\bibfnamefont {J.~G.}\ \bibnamefont {Keizer}}, \bibinfo {author} {\bibfnamefont {P.}~\bibnamefont {Pakkiam}}, \bibinfo {author} {\bibfnamefont {D.}~\bibnamefont {Keith}}, \bibinfo {author} {\bibfnamefont {M.~G.}\ \bibnamefont {House}}, \bibinfo {author} {\bibfnamefont {E.}~\bibnamefont {Peretz}},\ and\ \bibinfo {author} {\bibfnamefont {M.~Y.}\ \bibnamefont {Simmons}},\ }\bibfield  {title} {\bibinfo {title} {Spin read-out in atomic qubits in an all-epitaxial three-dimensional transistor},\ }\href@noop {} {\bibfield  {journal} {\bibinfo  {journal} {Nature Nanotechnology}\ }\textbf {\bibinfo {volume} {14}},\ \bibinfo {pages} {137} (\bibinfo {year} {2019})}\BibitemShut {NoStop}%
\bibitem [{\citenamefont {M{\"u}ller}\ \emph {et~al.}(2019)\citenamefont {M{\"u}ller}, \citenamefont {Cole},\ and\ \citenamefont {Lisenfeld}}]{muller2019}%
  \BibitemOpen
  \bibfield  {author} {\bibinfo {author} {\bibfnamefont {C.}~\bibnamefont {M{\"u}ller}}, \bibinfo {author} {\bibfnamefont {J.~H.}\ \bibnamefont {Cole}},\ and\ \bibinfo {author} {\bibfnamefont {J.}~\bibnamefont {Lisenfeld}},\ }\bibfield  {title} {\bibinfo {title} {Towards understanding two-level-systems in amorphous solids: insights from quantum circuits},\ }\href@noop {} {\bibfield  {journal} {\bibinfo  {journal} {Reports on Progress in Physics}\ }\textbf {\bibinfo {volume} {82}},\ \bibinfo {pages} {124501} (\bibinfo {year} {2019})}\BibitemShut {NoStop}%
\bibitem [{\citenamefont {Martinis}\ \emph {et~al.}(2005)\citenamefont {Martinis}, \citenamefont {Cooper}, \citenamefont {McDermott}, \citenamefont {Steffen}, \citenamefont {Ansmann}, \citenamefont {Osborn}, \citenamefont {Cicak}, \citenamefont {Oh}, \citenamefont {Pappas}, \citenamefont {Simmonds} \emph {et~al.}}]{martinis2005}%
  \BibitemOpen
  \bibfield  {author} {\bibinfo {author} {\bibfnamefont {J.~M.}\ \bibnamefont {Martinis}}, \bibinfo {author} {\bibfnamefont {K.~B.}\ \bibnamefont {Cooper}}, \bibinfo {author} {\bibfnamefont {R.}~\bibnamefont {McDermott}}, \bibinfo {author} {\bibfnamefont {M.}~\bibnamefont {Steffen}}, \bibinfo {author} {\bibfnamefont {M.}~\bibnamefont {Ansmann}}, \bibinfo {author} {\bibfnamefont {K.}~\bibnamefont {Osborn}}, \bibinfo {author} {\bibfnamefont {K.}~\bibnamefont {Cicak}}, \bibinfo {author} {\bibfnamefont {S.}~\bibnamefont {Oh}}, \bibinfo {author} {\bibfnamefont {D.~P.}\ \bibnamefont {Pappas}}, \bibinfo {author} {\bibfnamefont {R.~W.}\ \bibnamefont {Simmonds}}, \emph {et~al.},\ }\bibfield  {title} {\bibinfo {title} {Decoherence in josephson qubits from dielectric loss},\ }\href@noop {} {\bibfield  {journal} {\bibinfo  {journal} {Physical Review Letters}\ }\textbf {\bibinfo {volume} {95}},\ \bibinfo {pages} {210503} (\bibinfo {year} {2005})}\BibitemShut {NoStop}%
\bibitem [{\citenamefont {Gardner}(1992)}]{GARDNER1992113}%
  \BibitemOpen
  \bibfield  {author} {\bibinfo {author} {\bibfnamefont {W.~A.}\ \bibnamefont {Gardner}},\ }\bibfield  {title} {\bibinfo {title} {A unifying view of coherence in signal processing},\ }\href {https://doi.org/https://doi.org/10.1016/0165-1684(92)90015-O} {\bibfield  {journal} {\bibinfo  {journal} {Signal Processing}\ }\textbf {\bibinfo {volume} {29}},\ \bibinfo {pages} {113} (\bibinfo {year} {1992})}\BibitemShut {NoStop}%
\bibitem [{\citenamefont {Malekpour}\ \emph {et~al.}(2018)\citenamefont {Malekpour}, \citenamefont {Gubner},\ and\ \citenamefont {Sethares}}]{MALEKPOUR20182932}%
  \BibitemOpen
  \bibfield  {author} {\bibinfo {author} {\bibfnamefont {S.}~\bibnamefont {Malekpour}}, \bibinfo {author} {\bibfnamefont {J.~A.}\ \bibnamefont {Gubner}},\ and\ \bibinfo {author} {\bibfnamefont {W.~A.}\ \bibnamefont {Sethares}},\ }\bibfield  {title} {\bibinfo {title} {Measures of generalized magnitude-squared coherence: Differences and similarities},\ }\href {https://doi.org/https://doi.org/10.1016/j.jfranklin.2018.01.014} {\bibfield  {journal} {\bibinfo  {journal} {Journal of the Franklin Institute}\ }\textbf {\bibinfo {volume} {355}},\ \bibinfo {pages} {2932} (\bibinfo {year} {2018})}\BibitemShut {NoStop}%
\bibitem [{\citenamefont {Tosi}\ \emph {et~al.}(2017)\citenamefont {Tosi}, \citenamefont {Mohiyaddin}, \citenamefont {Schmitt}, \citenamefont {Tenberg}, \citenamefont {Rahman}, \citenamefont {Klimeck},\ and\ \citenamefont {Morello}}]{tosi2017}%
  \BibitemOpen
  \bibfield  {author} {\bibinfo {author} {\bibfnamefont {G.}~\bibnamefont {Tosi}}, \bibinfo {author} {\bibfnamefont {F.~A.}\ \bibnamefont {Mohiyaddin}}, \bibinfo {author} {\bibfnamefont {V.}~\bibnamefont {Schmitt}}, \bibinfo {author} {\bibfnamefont {S.}~\bibnamefont {Tenberg}}, \bibinfo {author} {\bibfnamefont {R.}~\bibnamefont {Rahman}}, \bibinfo {author} {\bibfnamefont {G.}~\bibnamefont {Klimeck}},\ and\ \bibinfo {author} {\bibfnamefont {A.}~\bibnamefont {Morello}},\ }\bibfield  {title} {\bibinfo {title} {Silicon quantum processor with robust long-distance qubit couplings},\ }\href@noop {} {\bibfield  {journal} {\bibinfo  {journal} {Nature Communications}\ }\textbf {\bibinfo {volume} {8}},\ \bibinfo {pages} {450} (\bibinfo {year} {2017})}\BibitemShut {NoStop}%
\bibitem [{\citenamefont {Mi}\ \emph {et~al.}(2017)\citenamefont {Mi}, \citenamefont {Cady}, \citenamefont {Zajac}, \citenamefont {Deelman},\ and\ \citenamefont {Petta}}]{mi2017}%
  \BibitemOpen
  \bibfield  {author} {\bibinfo {author} {\bibfnamefont {X.}~\bibnamefont {Mi}}, \bibinfo {author} {\bibfnamefont {J.}~\bibnamefont {Cady}}, \bibinfo {author} {\bibfnamefont {D.}~\bibnamefont {Zajac}}, \bibinfo {author} {\bibfnamefont {P.}~\bibnamefont {Deelman}},\ and\ \bibinfo {author} {\bibfnamefont {J.~R.}\ \bibnamefont {Petta}},\ }\bibfield  {title} {\bibinfo {title} {Strong coupling of a single electron in silicon to a microwave photon},\ }\href@noop {} {\bibfield  {journal} {\bibinfo  {journal} {Science}\ }\textbf {\bibinfo {volume} {355}},\ \bibinfo {pages} {156} (\bibinfo {year} {2017})}\BibitemShut {NoStop}%
\bibitem [{\citenamefont {Wang}\ \emph {et~al.}(2023)\citenamefont {Wang}, \citenamefont {Feng}, \citenamefont {Serrano}, \citenamefont {Gilbert}, \citenamefont {Leon}, \citenamefont {Tanttu}, \citenamefont {Mai}, \citenamefont {Liang}, \citenamefont {Huang}, \citenamefont {Su} \emph {et~al.}}]{wang2023}%
  \BibitemOpen
  \bibfield  {author} {\bibinfo {author} {\bibfnamefont {Z.}~\bibnamefont {Wang}}, \bibinfo {author} {\bibfnamefont {M.}~\bibnamefont {Feng}}, \bibinfo {author} {\bibfnamefont {S.}~\bibnamefont {Serrano}}, \bibinfo {author} {\bibfnamefont {W.}~\bibnamefont {Gilbert}}, \bibinfo {author} {\bibfnamefont {R.~C.}\ \bibnamefont {Leon}}, \bibinfo {author} {\bibfnamefont {T.}~\bibnamefont {Tanttu}}, \bibinfo {author} {\bibfnamefont {P.}~\bibnamefont {Mai}}, \bibinfo {author} {\bibfnamefont {D.}~\bibnamefont {Liang}}, \bibinfo {author} {\bibfnamefont {J.~Y.}\ \bibnamefont {Huang}}, \bibinfo {author} {\bibfnamefont {Y.}~\bibnamefont {Su}}, \emph {et~al.},\ }\bibfield  {title} {\bibinfo {title} {Jellybean quantum dots in silicon for qubit coupling and on-chip quantum chemistry (adv. mater. 19/2023)},\ }\href@noop {} {\bibfield  {journal} {\bibinfo  {journal} {Advanced Materials}\ }\textbf {\bibinfo {volume} {35}},\ \bibinfo {pages} {2370133} (\bibinfo {year} {2023})}\BibitemShut {NoStop}%
\bibitem [{\citenamefont {Keith}\ \emph {et~al.}(2019)\citenamefont {Keith}, \citenamefont {House}, \citenamefont {Donnelly}, \citenamefont {Watson}, \citenamefont {Weber},\ and\ \citenamefont {Simmons}}]{keith2019}%
  \BibitemOpen
  \bibfield  {author} {\bibinfo {author} {\bibfnamefont {D.}~\bibnamefont {Keith}}, \bibinfo {author} {\bibfnamefont {M.}~\bibnamefont {House}}, \bibinfo {author} {\bibfnamefont {M.}~\bibnamefont {Donnelly}}, \bibinfo {author} {\bibfnamefont {T.}~\bibnamefont {Watson}}, \bibinfo {author} {\bibfnamefont {B.}~\bibnamefont {Weber}},\ and\ \bibinfo {author} {\bibfnamefont {M.}~\bibnamefont {Simmons}},\ }\bibfield  {title} {\bibinfo {title} {Single-shot spin readout in semiconductors near the shot-noise sensitivity limit},\ }\href@noop {} {\bibfield  {journal} {\bibinfo  {journal} {Physical Review X}\ }\textbf {\bibinfo {volume} {9}},\ \bibinfo {pages} {041003} (\bibinfo {year} {2019})}\BibitemShut {NoStop}%
\bibitem [{\citenamefont {Strikis}\ \emph {et~al.}(2021)\citenamefont {Strikis}, \citenamefont {Qin}, \citenamefont {Chen}, \citenamefont {Benjamin},\ and\ \citenamefont {Li}}]{strikis2021}%
  \BibitemOpen
  \bibfield  {author} {\bibinfo {author} {\bibfnamefont {A.}~\bibnamefont {Strikis}}, \bibinfo {author} {\bibfnamefont {D.}~\bibnamefont {Qin}}, \bibinfo {author} {\bibfnamefont {Y.}~\bibnamefont {Chen}}, \bibinfo {author} {\bibfnamefont {S.~C.}\ \bibnamefont {Benjamin}},\ and\ \bibinfo {author} {\bibfnamefont {Y.}~\bibnamefont {Li}},\ }\bibfield  {title} {\bibinfo {title} {Learning-based quantum error mitigation},\ }\href@noop {} {\bibfield  {journal} {\bibinfo  {journal} {PRX Quantum}\ }\textbf {\bibinfo {volume} {2}},\ \bibinfo {pages} {040330} (\bibinfo {year} {2021})}\BibitemShut {NoStop}%
\bibitem [{\citenamefont {Temme}\ \emph {et~al.}(2017)\citenamefont {Temme}, \citenamefont {Bravyi},\ and\ \citenamefont {Gambetta}}]{temme2017}%
  \BibitemOpen
  \bibfield  {author} {\bibinfo {author} {\bibfnamefont {K.}~\bibnamefont {Temme}}, \bibinfo {author} {\bibfnamefont {S.}~\bibnamefont {Bravyi}},\ and\ \bibinfo {author} {\bibfnamefont {J.~M.}\ \bibnamefont {Gambetta}},\ }\bibfield  {title} {\bibinfo {title} {Error mitigation for short-depth quantum circuits},\ }\href@noop {} {\bibfield  {journal} {\bibinfo  {journal} {Physical Review Letters}\ }\textbf {\bibinfo {volume} {119}},\ \bibinfo {pages} {180509} (\bibinfo {year} {2017})}\BibitemShut {NoStop}%
\bibitem [{\citenamefont {Bravyi}\ \emph {et~al.}(2021)\citenamefont {Bravyi}, \citenamefont {Sheldon}, \citenamefont {Kandala}, \citenamefont {Mckay},\ and\ \citenamefont {Gambetta}}]{bravyi2021}%
  \BibitemOpen
  \bibfield  {author} {\bibinfo {author} {\bibfnamefont {S.}~\bibnamefont {Bravyi}}, \bibinfo {author} {\bibfnamefont {S.}~\bibnamefont {Sheldon}}, \bibinfo {author} {\bibfnamefont {A.}~\bibnamefont {Kandala}}, \bibinfo {author} {\bibfnamefont {D.~C.}\ \bibnamefont {Mckay}},\ and\ \bibinfo {author} {\bibfnamefont {J.~M.}\ \bibnamefont {Gambetta}},\ }\bibfield  {title} {\bibinfo {title} {Mitigating measurement errors in multiqubit experiments},\ }\href@noop {} {\bibfield  {journal} {\bibinfo  {journal} {Physical Review A}\ }\textbf {\bibinfo {volume} {103}},\ \bibinfo {pages} {042605} (\bibinfo {year} {2021})}\BibitemShut {NoStop}%
\bibitem [{\citenamefont {Hsueh}\ \emph {et~al.}(2014)\citenamefont {Hsueh}, \citenamefont {B\"uch}, \citenamefont {Tan}, \citenamefont {Wang}, \citenamefont {Hollenberg}, \citenamefont {Klimeck}, \citenamefont {Simmons},\ and\ \citenamefont {Rahman}}]{Hsueh2014}%
  \BibitemOpen
  \bibfield  {author} {\bibinfo {author} {\bibfnamefont {Y.-L.}\ \bibnamefont {Hsueh}}, \bibinfo {author} {\bibfnamefont {H.}~\bibnamefont {B\"uch}}, \bibinfo {author} {\bibfnamefont {Y.}~\bibnamefont {Tan}}, \bibinfo {author} {\bibfnamefont {Y.}~\bibnamefont {Wang}}, \bibinfo {author} {\bibfnamefont {L.~C.~L.}\ \bibnamefont {Hollenberg}}, \bibinfo {author} {\bibfnamefont {G.}~\bibnamefont {Klimeck}}, \bibinfo {author} {\bibfnamefont {M.~Y.}\ \bibnamefont {Simmons}},\ and\ \bibinfo {author} {\bibfnamefont {R.}~\bibnamefont {Rahman}},\ }\bibfield  {title} {\bibinfo {title} {Spin-lattice relaxation times of single donors and donor clusters in silicon},\ }\href {https://doi.org/10.1103/PhysRevLett.113.246406} {\bibfield  {journal} {\bibinfo  {journal} {Phys. Rev. Lett.}\ }\textbf {\bibinfo {volume} {113}},\ \bibinfo {pages} {246406} (\bibinfo {year} {2014})}\BibitemShut {NoStop}%
\bibitem [{\citenamefont {Stavrou}\ and\ \citenamefont {Hu}(2005)}]{Hu2005}%
  \BibitemOpen
  \bibfield  {author} {\bibinfo {author} {\bibfnamefont {V.~N.}\ \bibnamefont {Stavrou}}\ and\ \bibinfo {author} {\bibfnamefont {X.}~\bibnamefont {Hu}},\ }\bibfield  {title} {\bibinfo {title} {Charge decoherence in laterally coupled quantum dots due to electron-phonon interactions},\ }\href {https://doi.org/10.1103/PhysRevB.72.075362} {\bibfield  {journal} {\bibinfo  {journal} {Phys. Rev. B}\ }\textbf {\bibinfo {volume} {72}},\ \bibinfo {pages} {075362} (\bibinfo {year} {2005})}\BibitemShut {NoStop}%
\bibitem [{\citenamefont {Petit}\ \emph {et~al.}(2018)\citenamefont {Petit}, \citenamefont {Boter}, \citenamefont {Eenink}, \citenamefont {Droulers}, \citenamefont {Tagliaferri}, \citenamefont {Li}, \citenamefont {Franke}, \citenamefont {Singh}, \citenamefont {Clarke}, \citenamefont {Schouten}, \citenamefont {Dobrovitski}, \citenamefont {Vandersypen},\ and\ \citenamefont {Veldhorst}}]{Petit2018}%
  \BibitemOpen
  \bibfield  {author} {\bibinfo {author} {\bibfnamefont {L.}~\bibnamefont {Petit}}, \bibinfo {author} {\bibfnamefont {J.~M.}\ \bibnamefont {Boter}}, \bibinfo {author} {\bibfnamefont {H.~G.~J.}\ \bibnamefont {Eenink}}, \bibinfo {author} {\bibfnamefont {G.}~\bibnamefont {Droulers}}, \bibinfo {author} {\bibfnamefont {M.~L.~V.}\ \bibnamefont {Tagliaferri}}, \bibinfo {author} {\bibfnamefont {R.}~\bibnamefont {Li}}, \bibinfo {author} {\bibfnamefont {D.~P.}\ \bibnamefont {Franke}}, \bibinfo {author} {\bibfnamefont {K.~J.}\ \bibnamefont {Singh}}, \bibinfo {author} {\bibfnamefont {J.~S.}\ \bibnamefont {Clarke}}, \bibinfo {author} {\bibfnamefont {R.~N.}\ \bibnamefont {Schouten}}, \bibinfo {author} {\bibfnamefont {V.~V.}\ \bibnamefont {Dobrovitski}}, \bibinfo {author} {\bibfnamefont {L.~M.~K.}\ \bibnamefont {Vandersypen}},\ and\ \bibinfo {author} {\bibfnamefont {M.}~\bibnamefont {Veldhorst}},\ }\bibfield  {title} {\bibinfo {title} {Spin lifetime and charge noise in hot silicon quantum dot qubits},\ }\href
  {https://doi.org/10.1103/PhysRevLett.121.076801} {\bibfield  {journal} {\bibinfo  {journal} {Phys. Rev. Lett.}\ }\textbf {\bibinfo {volume} {121}},\ \bibinfo {pages} {076801} (\bibinfo {year} {2018})}\BibitemShut {NoStop}%
\bibitem [{\citenamefont {Aloisio}\ \emph {et~al.}(2023)\citenamefont {Aloisio}, \citenamefont {White}, \citenamefont {Hill},\ and\ \citenamefont {Modi}}]{Aloisio2023}%
  \BibitemOpen
  \bibfield  {author} {\bibinfo {author} {\bibfnamefont {I.~A.}\ \bibnamefont {Aloisio}}, \bibinfo {author} {\bibfnamefont {G.~A.}\ \bibnamefont {White}}, \bibinfo {author} {\bibfnamefont {C.~D.}\ \bibnamefont {Hill}},\ and\ \bibinfo {author} {\bibfnamefont {K.}~\bibnamefont {Modi}},\ }\bibfield  {title} {\bibinfo {title} {Sampling complexity of open quantum systems},\ }\href@noop {} {\bibfield  {journal} {\bibinfo  {journal} {PRX Quantum}\ }\textbf {\bibinfo {volume} {4}},\ \bibinfo {pages} {020310} (\bibinfo {year} {2023})}\BibitemShut {NoStop}%
\bibitem [{\citenamefont {Kattem{\"o}lle}\ and\ \citenamefont {Burkard}(2023)}]{Kattemolle2023}%
  \BibitemOpen
  \bibfield  {author} {\bibinfo {author} {\bibfnamefont {J.}~\bibnamefont {Kattem{\"o}lle}}\ and\ \bibinfo {author} {\bibfnamefont {G.}~\bibnamefont {Burkard}},\ }\bibfield  {title} {\bibinfo {title} {Ability of error correlations to improve the performance of variational quantum algorithms},\ }\href@noop {} {\bibfield  {journal} {\bibinfo  {journal} {Physical Review A}\ }\textbf {\bibinfo {volume} {107}},\ \bibinfo {pages} {042426} (\bibinfo {year} {2023})}\BibitemShut {NoStop}%
\bibitem [{\citenamefont {O'Brien}\ \emph {et~al.}(2001)\citenamefont {O'Brien}, \citenamefont {Schofield}, \citenamefont {Simmons}, \citenamefont {Clark}, \citenamefont {Dzurak}, \citenamefont {Kane}, \citenamefont {McAlpine}, \citenamefont {Hawley},\ and\ \citenamefont {Brown}}]{OBrien2001}%
  \BibitemOpen
  \bibfield  {author} {\bibinfo {author} {\bibfnamefont {J.~L.}\ \bibnamefont {O'Brien}}, \bibinfo {author} {\bibfnamefont {S.~R.}\ \bibnamefont {Schofield}}, \bibinfo {author} {\bibfnamefont {M.~Y.}\ \bibnamefont {Simmons}}, \bibinfo {author} {\bibfnamefont {R.~G.}\ \bibnamefont {Clark}}, \bibinfo {author} {\bibfnamefont {A.~S.}\ \bibnamefont {Dzurak}}, \bibinfo {author} {\bibfnamefont {B.~E.}\ \bibnamefont {Kane}}, \bibinfo {author} {\bibfnamefont {N.~S.}\ \bibnamefont {McAlpine}}, \bibinfo {author} {\bibfnamefont {M.~E.}\ \bibnamefont {Hawley}},\ and\ \bibinfo {author} {\bibfnamefont {G.~W.}\ \bibnamefont {Brown}},\ }\bibfield  {title} {\bibinfo {title} {{Towards the fabrication of phosphorus qubits for a silicon quantum computer}},\ }\href {https://doi.org/10.1103/PhysRevB.64.161401} {\bibfield  {journal} {\bibinfo  {journal} {Physical Review B - Condensed Matter and Materials Physics}\ }\textbf {\bibinfo {volume} {64}},\ \bibinfo {pages} {1614011} (\bibinfo {year} {2001})}\BibitemShut {NoStop}%
\bibitem [{\citenamefont {Simmons}\ \emph {et~al.}(2003)\citenamefont {Simmons}, \citenamefont {Schofield}, \citenamefont {O'Brien}, \citenamefont {Curson}, \citenamefont {Oberbeck}, \citenamefont {Hallam},\ and\ \citenamefont {Clark}}]{Simmons2003}%
  \BibitemOpen
  \bibfield  {author} {\bibinfo {author} {\bibfnamefont {M.~Y.}\ \bibnamefont {Simmons}}, \bibinfo {author} {\bibfnamefont {S.~R.}\ \bibnamefont {Schofield}}, \bibinfo {author} {\bibfnamefont {J.~L.}\ \bibnamefont {O'Brien}}, \bibinfo {author} {\bibfnamefont {N.~J.}\ \bibnamefont {Curson}}, \bibinfo {author} {\bibfnamefont {L.}~\bibnamefont {Oberbeck}}, \bibinfo {author} {\bibfnamefont {T.}~\bibnamefont {Hallam}},\ and\ \bibinfo {author} {\bibfnamefont {R.~G.}\ \bibnamefont {Clark}},\ }\bibfield  {title} {\bibinfo {title} {{Towards the atomic-scale fabrication of a silicon-based solid state quantum computer}},\ }\href {https://doi.org/10.1016/S0039-6028(03)00485-0} {\bibfield  {journal} {\bibinfo  {journal} {Surface Science}\ }\textbf {\bibinfo {volume} {532-535}},\ \bibinfo {pages} {1209} (\bibinfo {year} {2003})}\BibitemShut {NoStop}%
\bibitem [{\citenamefont {Shen}\ \emph {et~al.}(1995)\citenamefont {Shen}, \citenamefont {Wang}, \citenamefont {Abeln}, \citenamefont {Tucker}, \citenamefont {Lyding}, \citenamefont {Avouris},\ and\ \citenamefont {Walkup}}]{Shen1995}%
  \BibitemOpen
  \bibfield  {author} {\bibinfo {author} {\bibfnamefont {T.~C.}\ \bibnamefont {Shen}}, \bibinfo {author} {\bibfnamefont {C.}~\bibnamefont {Wang}}, \bibinfo {author} {\bibfnamefont {G.~C.}\ \bibnamefont {Abeln}}, \bibinfo {author} {\bibfnamefont {J.~R.}\ \bibnamefont {Tucker}}, \bibinfo {author} {\bibfnamefont {J.~W.}\ \bibnamefont {Lyding}}, \bibinfo {author} {\bibfnamefont {P.}~\bibnamefont {Avouris}},\ and\ \bibinfo {author} {\bibfnamefont {R.~E.}\ \bibnamefont {Walkup}},\ }\bibfield  {title} {\bibinfo {title} {{Atomic-scale desorption through electronic and vibrational excitation mechanisms}},\ }\href {https://doi.org/10.1126/science.268.5217.1590} {\bibfield  {journal} {\bibinfo  {journal} {Science}\ }\textbf {\bibinfo {volume} {268}},\ \bibinfo {pages} {1590} (\bibinfo {year} {1995})}\BibitemShut {NoStop}%
\bibitem [{\citenamefont {Lyding}\ \emph {et~al.}(1994)\citenamefont {Lyding}, \citenamefont {Shen}, \citenamefont {Hubacek}, \citenamefont {Tucker},\ and\ \citenamefont {Abeln}}]{Lyding1994}%
  \BibitemOpen
  \bibfield  {author} {\bibinfo {author} {\bibfnamefont {J.~W.}\ \bibnamefont {Lyding}}, \bibinfo {author} {\bibfnamefont {T.~C.}\ \bibnamefont {Shen}}, \bibinfo {author} {\bibfnamefont {J.~S.}\ \bibnamefont {Hubacek}}, \bibinfo {author} {\bibfnamefont {J.~R.}\ \bibnamefont {Tucker}},\ and\ \bibinfo {author} {\bibfnamefont {G.~C.}\ \bibnamefont {Abeln}},\ }\bibfield  {title} {\bibinfo {title} {{Nanoscale patterning and oxidation of H-passivated Si(100)-2x1 surfaces with an ultrahigh vacuum scanning tunneling microscope}},\ }\href {https://doi.org/10.1063/1.111722} {\bibfield  {journal} {\bibinfo  {journal} {Applied Physics Letters}\ }\textbf {\bibinfo {volume} {64}},\ \bibinfo {pages} {2010} (\bibinfo {year} {1994})}\BibitemShut {NoStop}%
\end{thebibliography}%

\end{document}